\documentclass[11pt,preprint]{aastex}

\newcommand{\e}{\epsilon}

\newcommand{\Op}{\Omega^\prime}

\newcommand{\tp}{t^\prime}
\newcommand{\ep}{\epsilon^\prime}
\newcommand{\mup}{\mu^\prime}
\newcommand{\bg}{\beta_\Gamma}

\newcommand{\rasim}{\lower.5ex\hbox{$\; \buildrel \rightarrow \over\sim \;$}}
\newcommand{\psim}{\lower.5ex\hbox{$\; \buildrel \propto \over\sim \;$}}
\newcommand{\llesssim}{\lower.5ex\hbox{$\; \buildrel \ll \over\sim \;$}}
\newcommand{\imp}{\lower.5ex\hbox{$\; \buildrel \rightarrow \over 
{\Gamma\gg 1} \;$}}
\newcommand{\iq}{\lower.5ex\hbox{$\; \buildrel \rightarrow \over {a\ll 1} \;$}}

\shorttitle{Model for Blazar Variability}
\shortauthors{Dermer and Schlickeiser}

\begin{document}
\title{Transformation Properties of External Radiation Fields,\\
Energy-Loss Rates and Scattered Spectra, \\
 and a Model for Blazar Variability
}

\author{Charles D. Dermer$^*$ and Reinhard Schlickeiser$^\dagger$}

\affil{$^*$Naval Research Laboratory, Code 7653, Space Science Division, 
Washington, DC 20375-5352}

\affil{$^\dagger$Institut f\"ur Theoretische Physik, Lehrstuhl IV: 
Weltraum- und Astrophysik, \\ Ruhr-Universit\"at Bochum, 
D-44780 Bochum, Germany}


\begin{abstract}

We treat transformation properties of external radiation fields in the
proper frame of a plasma moving with constant speed. The specific
spectral energy densities of external isotropic and accretion-disk
radiation fields are derived in the comoving frame of relativistic
outflows, such as those thought to be found near black-hole jet and
gamma-ray burst sources. Nonthermal electrons and positrons
Compton-scatter this radiation field, and high-energy protons and ions
interact with this field through photomeson and photopair production.
We revisit the problem of the Compton-scattered spectrum associated
with an external accretion-disk radiation field, and clarify a past
treatment by the authors. Simple expressions for energy-loss rates and
Thomson-scattered spectra are given for ambient soft photon fields
consisting either of a surrounding external isotropic monochromatic
radiation field, or of an azimuthally symmetric, geometrically thin
accretion-disk radiation field. A model for blazar emission is
presented that displays a characteristic spectral and variability
behavior due to the presence of a direct accretion-disk component.
The disk component and distinct flaring behavior can be bright enough
to be detected from flat spectrum radio quasars with {\it
GLAST}. Spectral states of blazars are characterized by the
relative importance of the accretion-disk and scattered radiation
fields and, in the extended jet, by the accretion disk, inner jet, and
cosmic microwave background radiation fields.

\end{abstract}


\section{Introduction}

The discovery of intense medium-energy gamma radiation from over 60
blazar AGNs with the EGRET instrument on the {\it Compton Observatory}
\citep{har99} shows that nonthermal gamma-ray production is an
important dissipation mechanism of jet energy generated by black-hole
accretion.  In the standard model for blazars, nonthermal synchrotron
emission is radiated by electrons accelerated via first-order shock
Fermi processes to Lorentz factors $\gamma \llesssim 5\times
10^7/\sqrt{B({\rm G})}$, where $B$ is the comoving magnetic
field. These electrons will Compton-scatter all available radiation
fields, including the internal synchrotron field and the external
radiation fields intercepted by the jet. The intensities of the
external ambient radiation fields, transformed to the comoving jet
frame, are  generally required  to model the production spectra of
blazars. External photon fields include the cosmic microwave
background radiation (CMBR) field (e.g., \citet{bur74,tav00,ds93},
hereafter DS93), the accretion-disk radiation field
(\citet{dsm92}, DS93), the accretion-disk radiation field scattered by
surrounding gas and dust
\citep{sik94,dss97}, infrared emissions from  hot dust or a
molecular torus \citep{pb97,bla00,aps02}, reflected synchrotron
radiation \citep{gm96,bd98}, broad-line region atomic line 
radiation \citep{1998ApJ...492..173K}, etc.

Gamma rays in leptonic models of broadband blazar emission originate
from synchrotron self-Compton (SSC) \citep{mgc92,bm96,tav98} or
external Compton (EC) (e.g.,  \citet{mk89}, DS93, \citet{bms97};
see \citet{bot01} and \citet{sm01} for recent reviews) processes. In
hadronic models, secondary photopairs and photomesons are produced
when energetic protons and ions interact with ambient synchrotron
photons (e.g., \citet{mb92,man93}) and photons of the external field
\citep{bd99,ad01}. The spectral energy distribution of a blazar will
consist, in addition to nonthermal synchrotron and SSC emissions, of
secondary radiations that are a consequence of processes involving
external soft photon fields. These radiations include neutral
secondaries from hadrons, and the radiations produced when external
soft photon fields are scattered by directly accelerated lepton
primaries and lepton secondaries formed in photopair, photohadronic,
and electromagnetic cascade processes.  The intensities of the various
components depend on the properties of the external fields, the
properties of the relativistic outflows, and the time-dependent
spectral injection of ions and electrons into the outflow. Given these
conditions, the evolving particle and photon distributions and
emergent radiation spectra can be derived for a specific model.

 A significant modeling effort has been devoted to blazar
studies. Where the synchrotron and SSC components are the dominant
radiation processes, as appears to be the case in some X-ray bright BL
Lac objects such as Mrk 421 and Mrk 501, observations of correlated
X-ray/TeV loop patterns found in graphs of spectral index versus
intensity are explained through combined particle acceleration,
injection, and radiative loss effects \citep{krm98,lk00}. The spectral
energy distributions of BL Lac objects are successfully modelled with
synchrotron and SSC components from broken power-law electron
distributions with low energy cutoffs, as in Mrk 501
\citep{mk97,pia98}. Models for the spectral energy distributions of
flat spectrum quasars, including external Compton components, have
been presented for different spectral states by \citet{har01},
\citet{muk99}, and \citet{bot99}, and in papers cited in the first 
paragraph. 

More recently, \citet{sik01} have presented a model for blazar
variability with time-dependent injection into a relativistically
moving blob that contains nonthermal electrons which are subject to
radiative and adiabatic losses. The scattered radiation from an
external, quasi-isotropic radiation field is treated, and evolving
spectral energy distributions and light curves are calculated. No
direct accretion disk field is however treated, though the emission
sites could be within 0.1-1 pc of the black hole, where such
components can make a significant contribution
\citep{ds94}. We show that the addition of the accretion-disk
components produce light curves and correlated multifrequency
behaviors that are quite distinct from the behavior calculated by
\citet{sik01}. These distinct patterns are bright enought
 to be well detected with the upcoming {\it Gamma ray Large Area Space
Telescope (GLAST)} mission,\footnote{see http://glast.gsfc.nasa.gov}
thus constraining the location of the shocks in gamma-ray blazars and
the dominant spectral components.

In this paper, we revisit the transformation properties of the
external radiation field, and clarify a past treatment by the authors
(DS93).  Here we adopt a formulation in terms of intensity rather than,
as in DS93, photon number. The intensity formulation seems to offer a simpler
method for comparing multiwavelength data, though the photon number approach 
perhaps simplifies the formulation of the scattering, attenuation, and 
secondary production interactions.  

The jet plasma is assumed to be located along the
symmetry axis of a geometrically-thin accretion disk.  The intensity
of photons in the jet frame depends on whether the accretion disk is
optically thick or thin at large disk radii $R
\gg r$, where $r$ is the distance of the jet plasma from the black hole.
The treatment of DS93 used an optically-thin formulation with a
volume-energy generation rate appropriate to disk accretion in a
steady flow. DS93 assigned a blackbody temperature for the mean energy
of the disk radiation field at different annuli, whereas a wide
variety of models are possible.  The radiation fields originating from
radii $R \gg r$, though spectrally distinct between the optically-thin
and -thick cases, give minor effect on the Thomson energy losses and
scattered photon spectra.

The energy density of disk photons that are scattered to $\gamma$-ray
energies by the jet electrons is shown to be dominated by photon
production at $R \approx r$ in the near-field (NF) regime, and by the
total black-hole power approximated as a receding point source in the
far-field (FF) regime.  The accretion-disk photons scattered by gas
and dust surrounding the black hole can dominate the near- and
far-field disk fields at large gravitational radii ($r_g = 1.5\times
10^{14} M_9$ cm), depending on the relative transition radii where the
NF, FF, and scattered external radiation fields dominate (Appendix A).

The transformed energy spectrum of the thermal accretion disk in the
 comoving frame displays a low-energy steepening for photons which
 originate primarily from radii $R \gg r$.  This is an important
 effect in photohadronic calculations of very high energy secondary
 production, and is also important in $\gamma$-$\gamma$ transparency
 calculations \citep{ad02}. Other photon sources can make an
 additional contribution in this range.  Because the bulk of the
 black-hole accretion power is generated at small disk radii, the use
 of an optically-thin or optically-thick formulation at $R\gg r$ has
 negligible effect on the total scattered power for power-law electron
 injection with injection index $>2$.  This article presents basic
 equations to calculate transformed external radiation fields from
 geometrically-thin accretion disks in the limiting optically-thin and
 optically-thick regimes.

Expressions for Thomson-scattered spectra from nonthermal electrons
with arbitrary energy and isotropic comoving pitch-angle distributions
are derived for external isotropic and direct accretion-disk radiation
fields. The electron equation-of-motion is solved in the regime where
the total synchrotron and external Thomson losses from the disk and
quasi-isotropic external radiation fields dominate SSC losses.  The
evolving electron distribution is obtained through a continuity
equation treatment with radiative losses on stationary electron
injection over a range of radii, but here neglecting adiabatic losses.
This simplified blazar model shows a dramatic change of spectral
behavior if nonthermal particle acceleration continuously operates
within $\approx 10^2$-$10^4 r_g$ from the black-hole engine. In the
model studied here, a pivoting of spectral shape is expected at a few
GeV on time scales of days, with $\gamma$-ray spectral activity
correlating with monotonic increases at IR and X-ray energies. Delayed
radio activity is predicted as the emergent energized plasma blob
expands and becomes optically thin to synchrotron self-absorption on
the parsec scale, though this behavior is not modeled here.  Campaigns
employing GLAST accompanied by broadband monitoring at other
frequencies can record multifrequency spectral flaring data and test
the model.

Preliminaries of radiation theory and invariants are given in Sections
2 and 3, respectively.  Section 4 treats the transformation of
external radiation fields, specializing to the azimuthally symmetric
and extreme-relativistic ($\Gamma \gg 1$) cases. The comoving specific
spectral energy density for an optically-thick \citet{ss73} accretion
disk is derived in Section 5.  Accretion-disk modeling is briefly
mentioned (\S 5.1) before proceeding to derive the comoving energy
spectrum of the disk field (\S 5.2). The same problem for an
optically-thin disk is treated in Section 6, and comparison with the
approach of DS93 is made in Section 7. The properties of the specific
spectral energy density of the accretion-disk radiation field,
transformed to the comoving fluid frame of the jet, are examined in
Section 8.

Section 9 presents analytic forms for the energy-loss rates and
spectra from various radiation processes, including synchrotron
emission (\S 9.1), radiation from Thomson-scattered external isotropic
monochromatic (\S 9.2) and accretion-disk fields in the near-field (\S
9.3) and far-field (\S 9.4) limits, and SSC radiation (\S 9.5). A
simple model for blazar variability is presented in Section 10 in the
limit that synchrotron and Thomson-scattered external radiation fields
dominate the electron-energy loss rates. Model results for stationary
particle acceleration and injection taking place within $\sim
10^2$-$10^4 r_g$ from the black hole reveal a characteristic spectral
behavior that joint radio/submillimeter/IR/X-ray/$\gamma$-ray
campaigns can test.  Detectability of blazar flares with {\it
GLAST} is treated in Section 11, where multiwavelength light
curves of a model blazar flare are presented.

The Appendices provide more detail on the derivations of the results
in Section 9. In particular, Appendices A.6 and A.7 show that
different spectral states of blazars and extended jets are defined by
the simple relations in equations (\ref{rcomparison}) and (\ref{rexj})
which govern the relative imporatance of the accretion-disk,
scattered, and CMBR fields.  An important deficiency in the model is
the treatment of Klein-Nishina effects in Compton scattering, which
has been considered in the context of blazar physics by \citet{bms97},
\citet{gkm01} and \citet{da02}. An analytic approach to this physics
is given in Appendix D, including comparison with the paper by
\citet{gkm01}.

\section{Preliminaries}

We consider external radiation fields from optically-thick and
optically-thin accretion disks, following preliminaries on radiation
physics \citep{nt73,rl79}. Our notation is such that a quantity
$X_\epsilon$ has the full units of $X$, whereas differential
quantities are denoted by $X(x_1,x_2;y) = \partial X(y)/\partial
x_1\partial x_2$, where $y$ is a parameter.

The intensity
\begin{equation}
I_\e(\Omega) = {d{\cal E}\over dAdtd\e d\Omega}
\label{I_e}
\end{equation}
is defined such that $I_\e(\Omega)dAdtd\Omega$ is the infinitesimal
energy $d{\cal E}$ carried during infinitesimal time $dt$ by photons with
dimensionless energy $\e = h\nu/m_ec^2$ between $\e$ and $\e +d\e$
that passes through area element $dA$ oriented normal to the direction
of rays of light lying within the solid angle element $d\Omega$ about
the direction $\vec\Omega$. Except for polarization,
$I_\e(\Omega;{\vec x},t)$ gives a complete description of the
radiation field at location $\vec x$ and time $t$.

The quantity $dF_\e = \mu I_\e(\Omega) d\Omega$ is the differential
spectral energy flux (units of ergs cm$^{-2}$ s$^{-1}$ $\e^{-1}$) that
passes through an area element oriented at an angle $\mu = \cos\theta
= \hat n\cdot \vec\Omega$ to the rays within solid angle element
$d\Omega$ about the direction $\vec
\Omega$. The net flux $F_\e(\hat n) = \oint d\Omega\;\mu I_\e
(\Omega)$ depends on the
orientation of the area element. Radiation fields that are azimuthally
symmetric about some axis and that can be expanded in even powers of
$\mu$ have zero net flux, that is, $F_\e = 0$. Obviously, the net flux
of an isotropic radiation field is zero.

Because the momentum of a photon is $h\nu/c$, the component of
momentum flux along $\hat n$ is given by $p_\e$ (dynes
cm$^{-2}\e^{-1}$) = $c^{-1} \oint d\Omega\;\mu^2 I_\e(\Omega)$. One
factor of $\mu$ is from the projection of the area element, and the
second is from the component of momentum along $\hat n$. Although
$I_\e(\Omega)\geq 0$ for all $\Omega$, $p_\epsilon$ can be $< 0$ if
the net momentum flux is directed opposite to $\hat n$. The total
radiation pressure is $p = c^{-1}\int_0^\infty d\e \oint d\Omega
\;\mu^2 I_\e(\Omega)$.

The $\nu F_\nu$ energy flux (ergs cm$^{-2}$ s$^{-1}$) at frequency
$\nu = m_ec^2\epsilon/h$ is  $f_\e(\hat n) = \e \, F_\e = \e
\oint d\Omega\, \mu I_\e(\Omega)$. The bolometric energy flux
$\Phi(\hat n) = \int_0^\infty d\e \, \e^{-1} f_\e(\hat n) =
\int_0^\infty d\e \oint d\Omega \, \mu I_\e(\Omega)$. The 
specific spectral energy density $u(\e,\Omega) = d{\cal E}/dVd\e d\Omega$,
and the spectral energy density $u(\e ) = \oint d\Omega \,
u(\e,\Omega)$.  Note that $d{\cal E} = u(\e,\Omega)dVd\e d\Omega = u(\e
,\Omega)cdtdAd\e d\Omega = I_\e(\Omega)dAdtd\e d\Omega$, so that
\begin{equation}
u(\e,\Omega) = c^{-1} I_\e(\Omega)\;.
\label{uvsI}
\end{equation}

The mean intensity $\bar I_\e = ({4\pi})^{-1}\oint d\Omega
\;I_\e(\Omega) = c u(\e)/(4\pi)$. Thus $u(\e) = 4\pi \bar I_\e/c$. The
total radiation energy density $u = \int_0^\infty d\e \; u(\e ) = 4\pi
c^{-1}\int_0^\infty d\e \;\bar I_\e $. The radiation pressure for an
isotropic radiation field with (mean) intensity $\bar I_\e^{iso}$ is $p =
c^{-1}\int_0^\infty d\e \oint d\Omega \;\mu^2 \bar I_\e^{iso} = (4\pi/3c)
\int_0^\infty d\e \;\bar I_e^{iso}$.  Because $u = (4\pi/c)\int_0^\infty d\e
 \; \bar I_\e^{iso}$, we have
\begin{equation}
p = u/3\;
\label{p=u}
\end{equation}
for an isotropic radiation field.

From conservation of energy, we can say that $d{\cal E}_1 = I_{\e,1}dA_1 dt
d\Omega_1 d\e_1 = I_{\e,2}dA_2 dt d\Omega_2 d\e_2$ by following a
bundle of rays from one area element to the next. The relationships
between area elements separated by distance $d$, and the solid angles
subtended by the bundle of rays at the two locations are $d\Omega_1 =
dA_2/d^2$ and $d\Omega_2 = dA_1/d^2$. With $\e_1 = \e_2 =\e$, $d\e_1 =
d\e_2 = d\e$, we obtain
\begin{equation}
I_{\e,1} = I_{\e,2}\;,\;\;{\rm or}\,\, dI_\e/ds =0\;,\;
\label{I(e)}
\end{equation}
provided no absorption occurs as the light rays propagate. The
equation of radiative transfer is
\begin{equation}
{dI_\e\over ds} = -\kappa_\e I_\e + j(\e,\Omega) \;\;,\;\;
 {\rm or}\;\;\; {dI_\e\over d\tau_\e} = - I_\e + S_\e ,
\label{dIds}
\end{equation}
where we define the photon-energy dependent spectral optical depth
$\tau_\e = \kappa_\e ds$ in terms of the $\e$-dependent spectral
absorption coefficient $\kappa_\e$ (cm$^{-1}$) and the emissivity
$j(\e,\Omega) = d{\cal E}/dVdtd\e d\Omega$.  The source function is
\begin{equation}
S_\e = {j(\e,\Omega)\over \kappa_\e}\;.
\label{S_e}
\end{equation}

The energy flux from a uniform brightness sphere of radius $R$ is an
instructive example. In this case, $I_\e(\Omega) = B_\e$, where $B_\e$
is independent of $\Omega$ by definition. Further defining
$\int_0^\infty d\e \, B_\e = B$, we find that the total energy flux at
the telescope aperture located a distance $d$ from the source is
$\Phi(\hat n) =
\int_0^\infty d\e \oint d\Omega \;\mu B_\e = 2\pi
B\int_{\sqrt{1-R^2/r^2}}^1 \;\mu\;d\mu  = \pi B R^2/r^2$.  Thus the
energy flux at the surface of a uniform brightness sphere is $\pi
B$. The energy flux at the surface of a blackbody radiator is
$\sigma_{\rm SB}T^4$, where the Stefan-Boltzmann constant $\sigma_{\rm
SB} = 5.67\times 10^{-5}$ ergs s$^{-1}$ cm$^{-2}$ degrees$^{-4}$.

\section{Invariant Quantities}

The elementary invariants are the invariant 4-volume $d^3\vec x dt =
dV dt$, the quantity $d^3\vec p/E$, and the invariant phase volume
$d{\cal V}= d^3\vec xd^3\vec p$. Here $E$ is photon or particle
energy, $p$ is momentum, and $d^3\vec p = p^2 dp d\Omega$. The
invariance of $d^3\vec x dt$ and $d^3\vec p/E$ can be demonstrated by
calculating the Jacobian for transformations consisting of rotations
and boosts. The invariance of $d{\cal V}$ follows by noting that
$dt/E$ is the ratio of parallel 4-vectors (Blumenthal and Gould 1970).

Because the number N of particles or photons is invariant, 
\begin{equation}
{dN\over d{\cal V}} = {1\over p^2}\; {dN\over dV dp d\Omega}
={1\over (m_ec)^{3}\, \e^{2}}\;{dN\over dVd\e d\Omega}
= {1\over m_e^4 c^5\e^{3}}\,{dE\over dVd\e d\Omega}
= \e^{-3}\; {u(\e,\Omega)\over m_e^4 c^5} \;,
\label{PdN/dV}
\end{equation}
where the latter three expressions apply to photons ($E =
mc^2\epsilon$). Hence, $\e^{-3} u(\e,\Omega)$ and
$\e^{-3}I_\e(\Omega)$ are invariants.

The function 
\begin{equation}
E\;{dN\over d^3\vec x dt d^3\vec p} =
{1\over (m_e c)^3 \e^2 }\;{dE\over dV dt d\e d\Omega}
 ={1\over (m_ec)^3\e^2}\;j(\e,\Omega)\;\;
\label{dNd3xd3p}
\end{equation}
is invariant, where the last two expressions apply to photons, implying 
that $\e^{-2} j(\e,\Omega)$ is invariant. This and the invariance
of $\e^{-3}I_\e(\Omega)$ show from equation (\ref{S_e}) that
$\e\kappa_\e$ is invariant.

\section{Transformation of External Radiation Fields}

 The photon energy $\e$ and angle $\theta = \cos^{-1}\mu$ in the
 stationary frame of the accretion disk are given in terms of the
 comoving photon energy $\ep$ and direction cosine $\mu^\prime$ 
through the relations $\e = \Gamma\ep (1+\beta_\Gamma\mu^\prime
 )$, $\mu = (\mu^\prime +\beta_\Gamma)/(1+\beta_\Gamma \mu^\prime)$,
 and $\phi = \phi^\prime$.  From the invariance of
 $u(\e,\Omega)/\e^3$, we have
\begin{equation}
u^\prime(\ep,\Omega^\prime )= {u(\e,\Omega)\over \Gamma^3(1+\bg\mup )^3}\;,
\label{uprime}
\end{equation}
For simplicity, we assume azimuthal symmetry about the axis of the
boost to the comoving jet frame, moving with bulk Lorentz factor
$\Gamma$ with respect to the stationary black hole system. For an
isotropic external photon field, $u(\e,\mu) = {1\over 2}u(\e)$. Hence
$u^\prime(\ep,\mup )= u(\e)/[2\Gamma^3(1+\bg\mup )^3]$ in this case.

For an external isotropic monochromatic radiation field, $u(\e) =
u_{ext}\delta(\e -\bar\e)$. Thus the comoving energy density
\begin{equation}
u_{ext}^\prime = \int_0^\infty d\ep \int_{-1}^1 d\mup u^\prime(\ep,
\mup )
= {u_{ext}\over 2\Gamma^4}\,\int_{-1}^1 d\mup\, (1+\bg\mup)^{-4}
 = u_{ext} \Gamma^2 (1+{\bg^2\over 3})\;
\label{uextprime}
\end{equation}
\citep{ds94}. Following the
integration over energy, one sees that the angle-dependent specific
energy density $u^\prime(\mup )= u_{ext}/[2
\Gamma^{4}(1+\bg \mup )^{4}]$. The four powers of the Doppler factor 
$\delta = \Gamma(1+\bg\mup) = [\Gamma(1-\bg\mu)]^{-1}$ are due to two
powers of $\delta$ from the solid angle transformation, one from the
energy transformation, and one from increased density due to length
contraction. For $\Gamma \gg 1$, the specific energy density ranges in
value from $\sim 8\Gamma^4 u_{ext}$ at $\mup = -1$ to $\sim
u_{ext}/(32\Gamma^4)$ at $\mup = 1$. The function plummets in value
when $\mup \gtrsim -\bg$. Multiplying $u^\prime(\mup=-\bg) \cong
\Gamma^4 u_{ext}/2$ by the characteristic solid angle element $\delta
\Omega^\prime \sim
\pi\theta^{\prime 2}
\sim\pi/\Gamma^2$ gives $u^\prime_{ext} \cong (\pi/2)\Gamma^2 u_{ext}$,
which approximately recovers equation (\ref{uextprime}) when $\Gamma
\gg 1$.  The function $u^\prime(\ep,\mup) \cong 4\Gamma^2 u_{ext}
\delta(\ep -4\Gamma\bar\e/3)\delta(\mup+1)/3$ provides a useful approximation
in the limit $\Gamma \gg 1$, noting that
$u^\prime_{ext}/n^\prime_{ext} = (1+\bg^2/3)\Gamma\bar\e m_ec^2$. For
isotropic external photons, the hardest and most intense radiation is
directed opposite to the direction of motion of the jet plasma in the
comoving frame.

Now consider the transformation of an isotropic, azimuthally symmetric
 power-law radiation
field described by the function
\begin{equation}
u(\e,\mu) =  k_\alpha u_{ext} \e^{-\alpha}\;,\;{\rm for~}
\e_\ell < \e < \e_u\;.
\label{uisopl}
\end{equation}
The normalization $u_{ext} = \int_0^\infty d\e \int_{-1}^1 d\mu\,
u(\e,\mu)$ gives
\begin{equation}
k_\alpha = \cases{{(1-\alpha)\over 2(\e_u^{1-\alpha}-
 \e_\ell^{1-\alpha})} \; ,& for general $\alpha$ \cr\cr [2\ln(\e_u/
 \e_\ell)]^{-1}\; , & for $\alpha = 1$ .\cr}\;\;
\label{kalpha}
\end{equation}
Equations (\ref{uprime}) and (\ref{uisopl}) imply
\begin{equation}
u^\prime(\ep,\mup ) = { k_\alpha u_{ext} \e^{\prime-\alpha}\over
\Gamma^3 (1+\bg\mup )^{3+\alpha}}\;,\;{\rm for~} \e_\ell <
\Gamma\ep(1+\bg\mup ) < \e_u\;,
\label{uprime2}
\end{equation}
which is easily shown to be normalized to $u^\prime_{ext}$ given by
equation (\ref{uextprime}). 

\section{Specific Spectral Energy Density}

The specific spectral energy density $u(\e,\Omega;\vec x,t) = c^{-1}
I(\e,\Omega)$ from equation (\ref{uvsI}). The external field at arbitrary
locations can be obtained from the constancy of specific intensity, 
equation (\ref{I(e)}), if intervening absorption is negligible. This
is the starting point for the calculations of external radiation
fields, from which transformed fields can be obtained through equation
(\ref{uprime}).

We specialize to the calculation of the external radiation field
emitted by steady, azimuthally symmetric emitting regions, as might
correspond to thin accretion disks around supermassive black holes. 
We furthermore consider a geometry where the emitting surface of the 
disk is located on the symmetry plane, and assume that the disk
thickness $h(R)$ at radius $R$ is $\ll R$.  Thus
\begin{equation}
\mu = {r\over\sqrt{r^2+R^2}} = (1+R^2/r^2)^{-1/2}\;, \;\; R = r(\mu^{-2}-1)^{1/2}\;.
\label{muR}
\end{equation}

In the cool, optically-thick blackbody solution of \citet{ss73}, the
disk emission is approximated by a surface radiating at the blackbody
temperature associated with the local energy dissipation rate per unit
surface area, which is derived from considerations of viscous
dissipation of the gravitational potential energy of the accreting
material \citep{st83}. One finds that the integrated emission spectrum
measured far from the black hole varies $\propto \epsilon ^{1/3}$ up
to a maximum photon energy associated with the innermost stable orbit
of the accretion disk.  The optically-thick solution is unstable in
the inner region near a black hole due to secular density/cooling
instabilities \citep{le74} in some regimes of the Eddington ratio
\begin{equation}
\ell_{\rm Edd} = {\eta\dot m c^2\over L_{\rm Edd}}\;,
\label{ellEdd}
\end{equation}
where $\eta$ is the efficiency to transform accreted matter to
escaping radiant energy.  The Eddington luminosity $L_{\rm Edd}=
1.26\times 10^{47} M_9$ ergs s$^{-1}$, where the mass of the central
supermassive black hole is $M =10^9M_9 M_\odot$ and the black hole is
accreting mass at the rate $\dot m$ (gm s$^{-1}$).
 
The intensity of a blackbody is 
\begin{equation}
I^{bb}_\epsilon(\Theta) = {2 c (m_ec^2)\over \lambda_{\rm
 C}^3}\;{\e^3\over [\exp(\epsilon/\Theta)-1]}\rightarrow {2 c
 (m_ec^2)\over \lambda_{\rm C}^3}\; \cases{\Theta\epsilon^2\; ,\;
 \e\ll\Theta & Rayleigh-Jeans regime \cr\cr
 \epsilon^3\exp(-\epsilon/\Theta) \; , \; \e\gg\Theta & Wien regime
 \cr}\;\;,
\label{InuT}
\end{equation}
where $\Theta = k_{\rm B}T/m_ec^2$ and $\lambda_{\rm C} = h/m_ec =
2.426\times 10^{-10}$ cm is the electron Compton wavelength.  For
steady flows where the energy is derived from the viscous dissipation
of the gravitational potential energy of the accreting matter, the
radiant surface-energy flux
\begin{equation}
{dE\over dAdt}  = 
{3GM\dot m\over 8\pi R^3}\; \varphi(R)
\label{PhiE}
\end{equation}
\citep{ss73}, where 
\begin{equation}
\varphi(R) =[1-\beta_i(R_i/R)^{1/2}]\;,
\label{varphiSchwarzschild}
\end{equation}
 $\beta_i \cong 1$, and $R_i = 6GM/c^2$ for a Schwarzschild
metric. Integrating equation (\ref{PhiE}) over a two-sided disk gives
$\eta = 1/12$.  Assuming that the disk is as an optically-thick
blackbody radiator, the effective temperature of the disk can be
determined by equating equation (\ref{PhiE}) with the surface energy
flux $\sigma_{\rm SB}T^4(R)$.

\subsection{Disk Models}

 To calculate the scattered jet radiation spectrum from the
disk-jet system, it is essential to properly characterize the
accretion-disk geometry and emissivity. Observations suggest that the
radius separating a geometrically-thin outer accretion disk from an
optically-thin hot inner cloud or disk increases with decreasing
$\ell_{\rm Edd}$ in the range $\ell_{\rm Edd} \lesssim $0.1.  (see
\citet{lia98} for a review of accretion disks around galactic black
hole sources).  Within an advection-dominated accretion disk (ADAF)
scenario (e.g., \citet{esin,dim,qg00}), the radiant luminosity from an
accreting system declines markedly when $\ell_{\rm Edd}\ll 0.1$ due to
the advection of photons into the black hole and the convection of
angular momentum and mass outward due to convective instabilities in
the advection-dominated flows. Thus $\eta = \eta(\dot m)$ in eq.\
(\ref{ellEdd}), and the radiant luminosity follows a steeper
dependence than $L_{rad}\psim \ell_{\rm Edd}$ when $\ell_{\rm Edd}\ll
0.1$.  The radio-loud branch of accreting black holes seems to be
found in the convectively unstable, low Eddington luminosity regime.

In this paper, we consider only the simplest flat-disk spectrum, which
is assumed to be well-approximated by the Shakura/Sunyaev disk
spectrum.  Optically-thin emission from the inner disk may also be
present and can be treated according the formulation presented here,
but is not considered here. The ejection of relativistic jet
plasma probably occurs in the low-luminosity ($\ell_{\rm Edd}\lesssim 0.1$)
regime, and evolutionary considerations
\citep{bd02} are consistent with this inference. 

The flux from this disk model is given by
\begin{equation}
I_\e^{\rm SS}(\Omega;R) \cong \;{3GM\dot m\over 16\pi^2
R^3}\varphi(R)\; \delta[\e - {2.7 k_{\rm B}\over m_ec^2}\;T(R)]\;
\label{uoverc}
\end{equation}
\citep{st83}. Here we use a monochromatic approximation for the mean
photon energy, with $T(R) $= $[3GM\dot m\varphi(R)$ $/8\pi
R^3\sigma_{\rm SB}]^{1/4}$.  It is simple to calculate the transformed
radiation field in the comoving frame of relativistic plasma along the
symmetry axis of the disk with this expression.

The outer accretion disk could be optically thin to Thomson
scattering, as we show in Section 6. When this happens, it is no
longer acceptable to adopt an optically-thick scenario.  Most of the
energy is dissipated in the central regions, because $\Delta E \propto
\Delta R/R^2$. Reprocessed radiations can dominate viscous radiations
in the outer disk, especially for flared disks.  The intensity of the
outer disk can then be dominated by reprocessed UV and X/$\gamma$
radiation, or emission from a surrounding torus.  In any case, the
emission of the outer disk is unlikely to be given precisely by
equation (\ref{uoverc}), though it provides a useful functional form
for further study.

\subsection{Integrated Emission Spectrum for Blackbody Disk Model}

The spectral energy density $u(\e ) = c^{-1}\oint d\Omega
\;I_\e^{\rm SS}(\Omega;R)$ along the jet symmetry axis is evaluated for
the Shakura/Sunyaev optically thick blackbody disk model from equation
(\ref{uoverc}). We define $K \equiv 2.7k_{\rm B} (3GM\dot m/ 8\pi
\sigma_{\rm SB})^{1/4}/ m_ec^2 $. Hence
\begin{equation}
u_{\rm SS}(\e) = {3GM\dot m\over 8\pi c}\;\int_0^{\mu_{max}} d\mu \;
\;{\varphi(R)\over R^3}\;
\delta[\e-K\varphi^{1/4}(R)R^{-3/4}]
\label{ue1}
\end{equation}
Eq.\ (\ref{ue1}) can be solved analytically in the approximation
$\varphi(R) \cong 1$, which becomes accurate in the limit $\e \ll
K/R_i^{3/4}$, giving
\begin{equation}
u_{\rm SS}(\e) \cong {3GM\dot m r\over 8\pi c}\;\int_{R_i}^\infty dR
 \; \;{\delta(\e - K R^{-3/4})\over R^2(r^2+R^2)^{3/2}}\; = {GM\dot m
 r\over 2\pi K c}\;{(\e/ K)^{1/3}\over [r^2 +(K/\e)^{8/3}]^{3/2}}
\label{uphe}
\end{equation}
$$\rightarrow {GM\dot m\over 2\pi K c}\cases{\;r^{-2}({\e\over
K})^{1/3}\; ,& for $r^{-3/4} \ll {\e \over K} \ll R_{i}^{-3/4}$ \cr\cr
\;r({\e\over K})^{13/3}\; ,& for ${\e\over K} \ll
r^{-3/4} $ \cr}\;.  $$ 
Thus we see that the rapid decline in the
energy radiated in the blackbody disk, compounded by the small solid
angle subtended by the area of the disk when $R\gg r$, leads to a 
spectral steepening in the integrated energy density of the external
radiation field.

Scaled quantities are adopted in order to get a better idea of the
photon energy of the external radiation field measured along the disk
axis for the blackbody disk model. The mass accretion rate $\dot m =
\ell_{\rm Edd} L_{\rm Edd}/(\eta c^2) = 1.4\times 10^{26}\ell_{\rm
Edd}M_9/\eta$ gm s$^{-1}$, using equation (\ref{ellEdd}). We write  $r
= \tilde r r_g$ and $R = \tilde R r_g$, where $\tilde r$ is the jet
height and $\tilde R$ is the disk radius in units of gravitational
radii $r_g = GM/c^2 = 1.48\times 10^{14}M_9$ cm. The characteristic
photon energy emitted from the disk at $\tilde R$ gravitational radii
from the nucleus is
\begin{equation}
m_ec^2\bar\e \cong 77 \;({\ell_{\rm Edd}\over M_9\eta})^{1/4}\tilde
R^{-3/4}\;\; {\rm eV}\;.
\label{mec2e}
\end{equation}
Replacing  the term $\tilde R$ in
equation (\ref{mec2e}) with the dimensionless jet height $\tilde r$ 
gives the photon energy at which the   integrated
spectral energy density $u_{\rm SS}(\e)$ 
displays a break from the $\e^{1/3}$ spectrum
to the $\e^{13/3}$ spectrum (eq.\ [\ref{uphe}]).  For a $10^9 M_\odot$
black hole accreting at the Eddington limit at efficency $\eta = 0.1$,
we therefore see that the mean photon energy radiated at 10, 10$^2$,
10$^3$ and $10^4$ gravitational radii from the central source is 26,
4.5, 0.81 and 0.14 eV, respectively. The photon energy derived at $10
r_g$ is, however, not that accurate because of the approximation
$\varphi \rightarrow 1$. Other effects such as gravitational
redshifting also become important when $\tilde R \lesssim 10$. In any
case, a modified blackbody or optically-thin disk model rather than a
blackbody model may hold in the inner disk region. Evidence for a
blackbody Shakura/Sunyaev disk around supermassive black holes is
provided by the intense optical/UV ``big blue bump" radiation, as
observed for example in the UV spectrum of 3C 273 \citep{lic95,kri99}.

\section{Transformation Properties of the Optically Thin Disk Radiation Field}

The continuity of mass in a steady flow consisting of ionized hydrogen
implies a mass accretion rate $\dot m = 2 \pi R h(R) v_{rad}(R) m_p
n(R)$, where $h(R)$ is the full disk thickness at radius $R$,
$v_{rad}(R)= c \beta_{rad}$ is the radial flow speed, and $n(R)$ is
the proton density at $R$. The vertical Thomson scattering depth
through the disk is
\begin{equation}
\tau_{\rm T} = \sigma_{\rm T} h(R) n(R) = {\sigma_{\rm T}
 \dot m\over 2\pi R c\beta_{rad} m_p}\;\cong 2.0\;{\ell_{\rm Edd}\over
\tilde R \beta_{rad}(\tilde R)\eta},
\label{tauT}
\end{equation}
If the middle-outer disk rotates in Keplerian motion, than
$mv_\theta^2 \sim GMm/R$ implies that $\beta_\theta\simeq \tilde
R^{-1/2}$, where $v_\theta =
\beta_\theta c$ is the azimuthal speed of the accretion flow. 
Letting $\beta_{rad} = k\beta_\theta$  ($k\lesssim 1$) implies $\tau_{\rm T} \cong
2\ell_{\rm Edd}/k \eta \tilde R^{1/2}$. If the radial flow speed is
sufficiently rapid, that is, if $\tilde R \gtrsim
 (2\ell_{\rm Edd}/k\eta)^2$, then the accretion disk is optically thin
to Thomson scattering. This condition requires $\eta \ll \ell_{\rm Edd}$,
which is compatible with ADAF models (\S 5.1). 

Within the Newtonian approximation for a thin accretion-disk geometry \citep{st83},
the surface energy flux is given by equation (\ref{PhiE}). If the
emitting region is optically thin to Thomson scattering, then
$dE/dVdtd\Omega = [2\pi h(R)]^{-1}(dE/dAdt)$ and the emissivity is
given by
\begin{equation}
j(\e,\Omega;R) = {3GM\dot m\over 16\pi^2 h(R)
R^3}\;\varphi(R) \delta[\e -\hat\e (R)]\; ,
\label{j(e,O)}
\end{equation}
where a monochromatic approximation to the emission spectrum is made.
The equation of radiative transfer (\ref{dIds}) for an optically-thin
region gives, noting that $\Delta s \cong h(R)/\mu$ for a thin disk,
\begin{equation}
I_\e(\Omega;R) = {3GM\dot m\over 16\pi^2 
R^3\mu}\;\varphi(R) \delta[\e -\hat\e (R)]\; .
\label{Ieothin}
\end{equation}

\section{Comparison with the Approach of Dermer and Schlickeiser (1993)}

The direct disk radiation field provides the most intense radiation
field to be scattered to gamma-ray energies when the relativistic
plasma ejecta is within some $10^3$-$10^4 r_g$ from the central source
\citep{ds94}, and even farther if the scattered radiation is weak. To
calculate the external radiation field intercepted by the ejecta, DS93
considered the emission function
\begin{equation}
n(\e,\Omega,R) = {\dot N(\e,R)\over 4\pi x^2 c}\; {\delta(\mu-
\bar\mu)\over 2\pi}\;,\; x = {r\over \mu}\;,
\label{neO}
\end{equation}
which is related to the specific spectral energy density according to
\begin{equation}
u(\e,\Omega) = m_ec^2 \int_{R_i}^\infty dR\; \e \;n(\e,\Omega,R)\;.
\label{uds}
\end{equation}
Equation (\ref{neO}) allows a large range
of radially-dependent emissivity functions to be tested, though the
formulation assumes optically-thin emissivity.

Note the relations
\begin{equation}
\e m_ec^2 \dot N(\e,R) = \dot E(\e,R) = 2\times 2\pi R \times {dE\over dA dt} 
\times \delta[\e-\hat\e(R)] =  {3GM\dot m\over 2  R^2}\;\varphi(R)\;
\delta[\e-\hat\e(R)].
\label{F(R)}
\end{equation}
Therefore
\begin{equation}
u_{\rm DS}(\e,\Omega) =\; {3GM\dot m\over 16\pi^2 c}\;\int_{R_i}^\infty dR\; 
{\delta(\mu - \bar\mu) \varphi(R)\;\delta[\e-\hat\e(R)]\over x^2 R^2}\;.
\label{FeR}
\end{equation} Transforming the $\delta$-function in $\mu$ to a $\delta$-
function in $R$ using equation (\ref{muR}) gives 
\begin{equation}
u_{\rm DS}(\e,\Omega) =  {3GM\dot m\over 16\pi^2 R^3 c\mu}\;\varphi(R)
\delta[\e - \hat\e(R)]\;,\;R = r\sqrt{\mu^{-2} -1},
\label{Ieds}
\end{equation}
equivalent to equation (\ref{Ieothin}).
By integrating equation (\ref{Ieds}) over $\Omega$ in the approximation
that $\varphi(R) \approx 1$, equation (\ref{uphe}) is recovered exactly when
$r^{-3/4} \ll \e /K \ll R_{i}^{-3/4}$. 
When $\e/ K \ll r^{-3/4} $, the exponent $13/3$ for the optically-thick result
becomes $3$ in the optically-thin formulation (compare eq.[\ref{uphe}]).

The approach of DS93 considers the external radiation field from an
optically-thin accretion disk that emits photons at a temperature or
energy corresponding to the blackbody value. Other optically-thin disk
models can be developed, including one- and two-temperature disk
models, or hybrid models, so we treated a specific case in our
paper. This description is valid to determine the transformation
properties of the disk when treated as an optically-thin radiator with
a blackbody disk spectrum, and differs only marginally from a
blackbody except when scattering photons with energies $\e \lesssim
Kr^{-3/4}$. The spectral effect makes little difference for Compton
scattering rates and spectra, but can be important in processes with
thresholds, e.g., photopion mechanism and $\gamma$-$\gamma$ absorption
\citep{ad02}.

A more general formulation must consider the transition between the
optically-thin and optically-thick regimes, geometrically-thick
accretion disks and more general disk models, and emergent jet
physics.

\section{Transformed Energy Density of the Accretion Disk Radiation Field}

As an illustrative example of the results presented in this paper, we
consider the optically-thick, geometrically-thin accretion-disk
radiation field (eq.\ [\ref{uoverc}]) in the approximation $\varphi
\rightarrow 1$, for which the specific spectral energy density along
the jet axis is
\begin{equation}
u(\epsilon,\Omega) = {3GM\dot m\over 16 \pi^2 R^3 c}\;\delta(\e-\hat\e)\;.
\label{ueo}
\end{equation}
The energy loss rate in the Thomson limit depends, for a distribution
of electrons with random pitch angle, only on the total comoving frame
energy density
\begin{equation}
u_{ext}^\prime = \int_{-\beta_\Gamma}^{\mu^\prime_{max}}d\mu^\prime
u^\prime(\mu^\prime) =
\oint d\Omega \int_0^\infty d\ep\; ({\ep\over \e})^3 u(\e , \Omega)\; = 
{3GM\dot m\over 8\pi c r^3 \Gamma } \;
\int_{-\beta_\Gamma}^{\mu^\prime_{max}}d\mu^\prime\; {(\mup +
\bg)^3\over (1-\mu^{\prime 2})^{3/2}(1+\bg\mup )^4}\;,
\label{uprimeext}
\end{equation}

Fig.\ 1 shows the integrand of the rightmost integral in equation
(\ref{uprimeext}).  When $\Gamma \gg 1$, two dominant components of
the differential energy density make up the total energy density: a
component from the disk radiation field at disk radii $R\approx r$,
called the {\it near-field} (NF) component; and a {\it far field} (FF)
component coming directly from behind, which dominates the disk
contribution at large radii \citep{ds93}.  The accretion disk
radiation field, when approximated as a point source that illuminates
the ejecta blob directly from behind, presents a total comoving energy
density
\begin{equation}
u^\prime_b = {1\over \Gamma^2(1+\bg^2)^2}\;{3GM\dot m\over 8\pi c r^2
R_i}\;(1-{2\over 3}\beta_i)\;.
\label{upb}
\end{equation}
 This result can be derived from the relation $u_b = 2\pi
 \int_0^{\mu_{max}}d\mu \int _0^\infty d\epsilon (I^{\rm
 SS}_\epsilon/c)$, using equations (\ref{uoverc}) and
 (\ref{varphiSchwarzschild}) in the limit $r \gg R$, combined with the
 relativistic transformation of the energy density (eq.[6] in
 \citet{ds94}).
The point-source approximation improves as the accretion disk looks
more like a point source in the comoving frame, that is, when
\begin{equation}
\mu^\prime_{max} \cong {1-(R_{min}^2/2r^2)-\bg \over
 1-\bg(1-R^2_{min}/2r^2)}\cong 1 - 2({\Gamma R_{min}\over r})^2\;\rightarrow 1.
\label{mup}
\end{equation}

The far-field approximation thus holds only when $r\gg \Gamma
R_{min}$,  and dominates the near-field component only when $r\gg
\Gamma^4 R_{min}$ (Appendix A.3). The following relationship
connects the disk model with the disk radiant luminosity, neglecting
advective effects:
\begin{equation}
L_d = 2\times2\pi \times \int_{R_{min}}^\infty dR\cdot R\cdot {dE\over
dAdt} = {3GM\dot m\over 2 R_i} (1- {2\beta_i\over 3}) \cong \;
3^{-\beta_i} \, ({GM\dot m\over 4 r_g})\; .
\label{Ldisk2}
\end{equation}
In Appendices A1 and A2, analytic properties are derived of the
integrand of the right-most term in equation (\ref{uprimeext}) in the
NF and FF limits, respectively.

\section{Analytic Energy-Loss Rates and Spectra}

This provides sufficient theory to derive expressions for radiation
spectra in blazar jets.  Simplified analytic expressions for
electron-energy loss rates and spectral components are presented in
the near-field and far-field regimes. For completeness, we also
summarize results of our previous work for synchrotron and synchrotron
self-Compton processes \citep{ds93,dss97}.  These expressions employ
$\delta$-function approximations for the emission spectra, and the
spectral forms are least accurate near endpoints and spectral breaks.

\subsection{Synchrotron Radiation}

The rate at which a randomly ordered pitch-angle distribution of
relativistic nonthermal electrons lose energy via the synchrotron
process in a region with mean comoving magnetic-field intensity $B$ is
given by $-m_ec^2\dot\gamma_{syn}$, where
\begin{equation}
-\dot\gamma_{syn} = {4\over 3}\,
{c\sigma_{\rm T}u_B\over m_ec^2}\,\gamma^2 \cong 1.3
\times 10^{-9}B^2\gamma^2\; {\rm s}^{-1} \;\equiv\; k_{syn}\gamma^2
\label{dotgammasyn}
\end{equation}
\citep{bg70}, and $u_B = B^2/8\pi$ is the energy density of the magnetic field.

In the $\delta$-function approximation for the elementary synchrotron
emissivity, the $\nu F_\nu$ synchrotron radiation spectrum from a
uniform blob, which is assumed to be spherical and to have a randomly
oriented magnetic field in the comoving frame, is
\begin{equation}
f_\epsilon^{syn} \simeq \delta^4\;({c\sigma_{\rm T} u_B\over 6\pi d_L^2})\;
\bar\gamma^3N_e(\bar \gamma)\;,\;\bar \gamma = \sqrt{{(1+z)
\epsilon\over \delta\epsilon_B}}\; .
\label{gdotsyn}
\end{equation} 
In this expression, emission properties are integrated over variations
on the comoving size scale of the plasma blob.  On the corresponding
observer time scale $t_{var} = (1+z)r_b/c\delta $, the blob upon
reaching location $x$ is assumed to host an instantaneous electron
energy spectrum $N_e(\gamma )$. Here $N_e(\gamma;x)d\gamma$ is the
differential number of electrons with comoving Lorentz factors between
$\gamma$ and $\gamma +$ d$\gamma$. $N_e(\gamma;x)$ is evaluated by
solving a continuity or diffusion equation. Better descriptions of the
system require integrations over the emitting volumes that arise from
light-travel time effects \citep{cg99,cd99,gps99}.

\subsection{Thomson-Scattered External Isotropic Monochromatic Radiation Field}

Consider the Thomson energy-loss rate of electrons that are randomly
 distributed throughout a plasma blob that passes through a uniform external
 isotropic monochromatic radiation field with mean photon energy
 $\epsilon_* = 10^{-4}\epsilon_{-4}$. The energy-loss rate due to
 Compton-scattered CMBR in the Thomson limit is given through
\begin{equation}
-\dot\gamma_{\rm T} = {4\over 3}\;{c\sigma_{\rm T}\over
m_ec^2}\gamma^2 u_*\Gamma^2(1+\beta_\Gamma^2/3)
\cong 3.3\times 10^{-8}u_*[{{\rm ~ergs~}\over{\rm cm}^{2}}]
\gamma^2\Gamma^2(1+\beta_\Gamma^2/3)
\;{\rm s}^{-1}\;\equiv k_{\rm T}\gamma^2\;,
\label{dotgammaT}
\end{equation} 
where $u_*$ is the  energy density of the external isotropic radiation field
photons which are scattered in the Thomson limit
\citep{ds93,ds94}. Equation (\ref{dotgammaT}) holds for $\gamma\ll
\gamma_{\rm KN}$, where $\gamma_{\rm KN}$ is the Lorentz factor where
Klein-Nishina (KN) effects become important, given through
$4\gamma_{\rm KN}\Gamma\epsilon_* = 1$.  The transition from Thomson
scattering to KN scattering takes place at observed photon energies
$\epsilon_{\rm KN} \cong \delta/[16\Gamma\epsilon_*(1+z)] \cong 300
(\delta/\Gamma)[\epsilon_{-4}(1+z)]^{-1}$ MeV (see
\citet{bms97,gkm01}, and \citet{da02} for treatments of KN effects on
external Compton scattering). 

The $\nu F_\nu$ spectrum radiated by relativistic electrons which
 scatter photons from an external isotropic monochromatic radiation
 field in the Thomson limit is
\begin{equation}
f_\epsilon^{\rm T} \simeq \delta^6\;{c\sigma_{\rm T} u_*\over (p+3)\pi
d_L^2}\;\bar\gamma^3N_e(\bar \gamma)\;,\;\bar \gamma =
\delta^{-1}\sqrt{{(1+z)\epsilon\over 2\epsilon_*}}\;
\label{gdotT}
\end{equation}
\citep{dss97}. The mean ambient CMB photon energy $\epsilon_* = $ 
$\hat\epsilon(1+z)$, $\hat\epsilon \cong 2.7k_{\rm B}T_{CMB}
/m_ec^2 = $ $1.24\times 10^{-9}$,  where $T_{CMB} = 2.72$ K is the 
present temperature of the CMB, 
and $u_* = 4\times 10^{-13}(1+z)^4$
ergs s$^{-1}$ is the local ambient CMBR energy density. For blazars, the
ambient photon field might be reprocessed UV accretion disk radiation,
with energy density depending on principle resonance line transitions
in the broad line region or illuminated disk and torus.  A simple
prescription for the scattered radiation field is to let
\begin{equation}
u_* =
{ L_d\tau_{sc}\over 4\pi r_{sc}^2c} = 3.5\times 10^{-4} \,{\ell_{\rm Edd}
M_9\over r_{sc}^2({\rm pc})}\;({\tau_{sc}\over 0.01 })\; ,\;  {\rm where~} 
\; L_d= \ell_{\rm Edd}L_{\rm Edd} = \eta \dot m
c^2 {\rm ~ergs}^{-1}\;,
\label{u*BLR}
\end{equation}
 $\tau_{sc}$ is an
effective scattering depth of the surrounding medium including the
broad line region, and $r_{sc} $ is the characteristic size of the
surrounding scattering medium \citep{ds94}. Effects of spatially varying external
radiation energy densities from scattering media with power-law
density gradients are
treated by \citet{bl95}. 

The distance along the jet axis beyond where the quasi-isotropic
scattered radiation field arising, for example, from disk radiation
scattered by broad emission-line clouds, dominates the point-source
disk radiation field is $\cong 0.04
R_{pc}\Gamma_{10}^{-2}(\tau_{sc}/0.01)^{-1/2}$ pc \citep{ds94}. The
appearance of weaker emission line fields in BL Lac objects suggests
that accretion-disk rather than scattered soft-photon radiation could
make a larger relative contribution to the SSC component in the inner
jets of BL Lac objects, and this component should be included in
detailed spectral models using more realistic accretion disk
models. The radii where Thomson losses from the external scattered
radiation field dominates the fields described in the NF and FF
regimes are derived in Appendices A4 and A5, respectively. Appendix A6
describes different spectral states of blazars according to the
dominant electron energy-loss processes. Appendix A7 considers the
transition radius where the electron energy-loss rates from the CMBR
field dominates those of the external accretion-disk radiation fields.

\subsection{Thomson-Scattered Near-Field Accretion Disk Radiation Spectrum}

Electrons lose energy when scattering soft photons of the
accretion-disk radiation field.  Considering an optically-thick
Shakura and Sunyaev accretion disk radiation spectrum, the electron
energy-loss rate in the Thomson limit of the near field ($r \approx R
\gg 10 r_g$) regime is given through
\begin{equation}
-\dot\gamma_{NF} \cong {4\over 3}\, c\sigma_{\rm T}\,[{0.7 L_d\over
4\pi r^2 c (m_ec^2)}\,({r_g\over r})]\Gamma^2\gamma^2\; \equiv \; 
k_{NF}\;{\gamma^2\over \tilde r^3}\;.
\label{Edotnf}
\end{equation}
	
The scattered $\nu F_\nu$ radiation spectrum for the near-field
component of the disk field is
\begin{equation}
f_\epsilon^{NF} \simeq {1\over 2}\,\delta^6\,{c\sigma_{\rm T}\over
6\pi d_L^2}\;({L_dr_g\over 4\pi r^3 c})\,\bar \gamma^3 N_e(\bar
\gamma)\;,\;\bar\gamma \cong {\sqrt{2}\over
\delta}\;\sqrt{{(1+z)\epsilon\over \bar\epsilon(\sqrt{3}r)}} \;.
\label{fnf}
\end{equation}
The derivation of these results is found in Appendix B.

\subsection{Thomson-Scattered Far-Field Accretion Disk Radiation Spectrum}

Again considering an optically-thick Shakura and Sunyaev accretion
disk radiation spectrum, the electron energy-loss rate in the Thomson
limit of the far-field regime is given through
\begin{equation}
-\dot\gamma_{FF} \cong {4\over 3}\, c\sigma_{\rm T}\,[{ L_d\over 4\pi
r^2 c (m_ec^2)}]\;{\gamma^2\over
\Gamma^2(1+\beta_\Gamma)^2}\; \equiv \; k_{FF}\;{\gamma^2\over \tilde r^2}\;.
\label{Edotff}
\end{equation}
	
The scattered $\nu F_\nu$ radiation spectrum for the far-field
component of the disk field is
\begin{equation}
f_\epsilon^{FF} \simeq \delta^6\,{c\sigma_{\rm T}\over 8\pi
d_L^2}\;(1-\mu^2)\;({L_d\over 4\pi r^2 c})\,\bar \gamma^3 N_e(\bar
\gamma)\;,\;\bar\gamma \cong {1\over
\delta}\;\sqrt{{(1+z)\epsilon\over(1-\mu) \bar\epsilon(10r_g)}} \;.
\label{fff}
\end{equation}
The derivation of these results is found in Appendix C.  By comparing
equations (\ref{Edotnf}) and (\ref{Edotff}), we see that the
transition from the dominance of the NF to the FF (point source)
behavior occurs at the transition altitude $r_{tr}\approx 3\Gamma^4
r_g$ (\citet{ds93}; see Appendix A3).

\subsection{Synchrotron Self-Compton Radiation}

The energy-loss rate of electrons as they Compton scatter their
self-synchrotron emission is, in the Thomson limit,
\begin{equation}
-\dot\gamma_{SSC,T}\simeq {4\over 3}\;{c\sigma_{\rm T}\over m_ec^2}\,
[\int_0^{1/\gamma} \; u^\prime(\ep)d\ep ]\;\gamma^2
\label{gammadotssct}
\end{equation}
The spectral energy density can be directly related to observables
through
\begin{equation}
\ep u^\prime(\ep ) \cong {2 d_L^2\over r_b^2 c}\; 
{f_\epsilon\over \delta^4}\; , \; \epsilon =
{\delta\ep \over (1+z)} \;{\rm and~}  r_b 
\sim {c\delta t_{var}\over 1+z}\;,
\label{epu}
\end{equation} 
where the last expression relates the
comoving size scale to the variability time scale and redshift through
$\delta$ \citep{tav98}.

The $\nu F_\nu$ SSC spectrum in the $\delta$-function approximation
for the synchrotron and Thomson emission spectra is
\begin{equation}
f_\e^{SSC} \simeq {3 \delta^4 c\sigma_{\rm T}^2 u_B
\over  36 \pi^2 r_b^2d_L^2}\; ({\epsilon^{\prime}\over \epsilon_B })^{3/2}\;
\int_{\epsilon_B}^{{\rm min}(\epsilon^\prime,1/\epsilon^{\prime})}
d\epsilon^{\prime\prime}\;
\epsilon^{\prime\prime -1}\,
N_e[({\epsilon^{\prime\prime}\over \epsilon_B})^{1/2}]
N_e[({\epsilon^{\prime}\over \epsilon^{\prime\prime}})^{1/2}]
\label{SSC}
\end{equation}
\citep{dss97}, where $\epsilon^\prime$ is related to $\epsilon$
in equation (\ref{epu}). Note that a factor $m_ec^2$ is missing 
in the calculation of the SSC spectrum in the
paper by \citet{dss97}, so that its importance is underestimated.
\citet{tav98} give expressions when the synchrotron 
spectrum is approximated by
a broken power law.

\section{Model for Blazar Variability}

We operate in the regime where Thomson and synchrotron losses dominate
SSC losses, adiabatic losses can be neglected, $B$ can be treated as
constant, and processes involving external fields can be treated for
the thermal accretion-disk model as considered in Section 9. The
analytic solution presented below can also be extended to the cases 
where $B^2 \propto r^{-2}$ and $B^2 \propto r^{-3}$, though we 
only consdier a constant magnetic field in the calculations 
presented here.  The
neglect of Klein-Nishina effects on the electron energy loss rate and
spectrum is a severe limitation of this model (see App.\ D). Accurate
calculations employing realistic spectral forms for synchrotron and
Compton processes are given in the papers by \citet{bms97},
\citet{bot99}, \citet{muk99}, and \citet{har01}.

In the stated approximations,
\begin{equation}
-\dot \gamma = [(k_{syn}+k_{T})+{k_{FF}\over \tilde r^2}+
{k_{NF}\over \tilde r^3}]\gamma^2\;.
\label{dotgammatot}
\end{equation}
Location is specified in dimensionless units of $\tilde r = r/r_g$,
$dt^\prime = r_g d\tilde r/(\bg\Gamma c)$, and the coefficients follow
from the results of the previous section. Solving equation
(\ref{dotgammatot}) gives
\begin{equation}
\gamma=\gamma(\gamma_i,\tilde r,\tilde r_i) = [\gamma_i^{-1}+F(\tilde r,\tilde r_i)]^{-1}\;,
\label{gamma}
\end{equation}
where
\begin{equation}
F(\tilde r,\tilde r_i)= {r_g\over \bg \Gamma
c}\;[(k_{syn}+k_{\rm T})(\tilde r -\tilde r_i) +k_{FF}({1\over \tilde
r_0} - {1\over \tilde r})+ {1\over 2}k_{NF}({1\over \tilde r^2_i} -
{1\over \tilde r^2})]\; ,
\label{Frri}
\end{equation}
where $r_0$ is the location at which the particle injection begins.

Within the framework of first-order Fermi acceleration theory, the
injection spectrum of particles downstream of the shock is
approximated by
\begin{equation}
{dN^\prime_e(\gamma_i)\over d\tp } = K
\gamma_i^{-p}H[\gamma_i;\gamma_{min},\gamma_{max}]\; ,\;{\rm and~} K =
{(p-2)L_e^\prime\over m_ec^2 (\gamma_{min}^{2-p} -\gamma_{max}^{2-p})}
\label{injection}
\end{equation}
where H[x;a,b] is a Heaviside function such that $H = 1$ for $a\leq x
 < b$ and $H = 0$ otherwise.  The term $\gamma_{max}$ is determined by
 size scale, radiation, acceleration, available-time and kinematic
 limits \citep{jag96,vie98,rm98,dh01}. The normalization is easily
 made to the injection power into nonthermal electrons $L_e^\prime =
 dE^\prime_e /d\tp = dE_*/dt_*$, where the invariance of injection
 power between the stationary frame of the black-hole jet (starred)
 system and the comoving (primed) system is invoked.  Hence the
 instantaneous electron spectrum at location $\tilde r$ is
\begin{equation}
N_e^\prime(\gamma;\tilde r) = {Kr_g\over \bg\Gamma
c\gamma^2}\;\int_{\tilde r_0}^{\min[\tilde r_1,\tilde r(t)]} \;d\tilde
r^\prime\;\gamma_i^{2-p}\, H[\{\gamma^{-1} - F(\tilde r^\prime,\tilde
r_i)\}^{-1};\gamma_{min},\gamma_{max}]\;.
\label{steady}
\end{equation}
 The term $r_1$ is the location where particle injection ends, and 
$r(t)$ is the location of the jet that emits radiation observed
at time $t$. 

Equation (\ref{steady}) is easily solved using equations (\ref{gamma}) and (\ref{Frri})
to give the instantaneous electron spectrum in the approximation that
$\Gamma$, $B$, and the injection power $L^\prime_e$ are constant with
time.  The allowed range of $\gamma$ is set by the Heaviside function and the limits
on the integral, implying $\gamma > [\gamma_{min}^{-1} +F(\tilde r,\tilde r_0)]^{-1}$ for
all values of $\tilde r \geq \tilde r_0$. When $\tilde r_0 \leq \tilde r_1$,
$\gamma < [\gamma_{max}^{-1} +F(\tilde r,\tilde r_0)]^{-1}$, whereas when $\tilde r \geq r_1$,
$\gamma < [\gamma_{max}^{-1} +F(\tilde r,\tilde r_1)]^{-1}$.
 \citet{mp00} were the first to modify the the approach
of \citet{ds93} to extended injection.
The methods of blast-wave physics \citep{mes02}
may be used to set $\gamma_{min}$ and $B$ and treat the spatial evolution
of $\Gamma$ due to interactions with an external medium, at least in
the adiabatic and fully radiative regimes. In this paper, we simply assign these
values.

Fig.\ 2 shows the evolution of the spectral components in the toy
model for a standard parameter set: $\Gamma = 20$, $\theta =
1/\Gamma$, $M_9 = 0.1$, $L_{\rm Edd} = 1$, $\tau_{sc} = 0.01$, $L_e =
10^{44}$ ergs s$^{-1}$, $t_{var} = 1$ day, $z=1$ ($d_L = 2.4\times
10^{28}$ cm for a cosmology with 70\% dark energy and a Hubble
constant of 65 km s$^{-1}$ Mpc$^{-1}$), $B = 1$ Gauss, $\gamma_{min} =
10^3$, $\gamma_{max} = 10^5$, $r_{sc} = 0.1$ pc, and $p = 2.3$.
Because the electron power is small in comparison with the Eddington
luminosity, the system could be accreting well below the Eddington
limit, as argued in an evolutionary scenario by
\citet{bd02}. Different disk models appropriate to smaller values of
the Eddington ratio can be treated in future work.

The injection is uniform between 1000 $r_g$ and 1500 $r_g$, 2000
$r_g$, 4000 $r_g$, 7000 $r_g$, and 10000 $r_g$ in Figs. 2a, 2b, 2c,
2d, and 2e, respectively.  The relation $dr = \bg \Gamma c \delta
dt/(1+z)$ implies
\begin{equation}
\tilde r_3 \equiv {r(t)\over 1000 r_g} = 0.8\; 
{t_4\over M_9 (1+z)} ({\Gamma\over 20})^2 \; ({\delta\over \Gamma})\;,
\label{r(t)}
\end{equation}
where $t_4$ is the observing time in units of $10^4$ s, measured from
the beginning of the flare when the emitting plasma was at $\tilde
r_0$. For our standard values with $M_9 = 0.1$ and $z=1$, a $10^4$ s
duration flare corresponds to the time during which the emitting blob
travelling $4000 r_g$.

This characteristic time scale
is optimum for examining flares with GLAST in terms of
flux levels for detecting intraday variability and proposed GLAST
slewing strategies, as discussed in the next section.  As can
be seen in Fig.\ 2,, a dominant spectral feature from the near-field accretion
disk component appears early in the flare, during which the $X/\gamma$
continuum is at a low lever. As the flare progresses, the near-field
component becomes increasingly weak, and the bulk of the gamma-ray
emission begins to originate from external scattered soft disk
photons. The $\approx 10$ MeV - 1 GeV, X-ray and synchrotron
radiations decline while the $> 1$ GeV radiation monotonically
increases. In addition to KN effects on the particle and photon
spectra, the diffuse intergalactic infrared radiation field will
attenuate $\gg 10 $ GeV radiation. The important cosmological
implications of this effect are discussed by, e.g., \citet{pri99} and
\citet{ss98}.

Fig.\ 3 shows the evolution of the SED with distance from the
black-hole core. For constant Lorentz factor $\Gamma = \delta = 20$
and $M_9 = 0.1$, the duration of the episode from $10^3$ to $10^4 r_g$
is 24 ksec.  The pivoting behavior in the GLAST energy range is
evident. Different injection profiles can change the the detailed
behavior, but the discovery of $\gamma$-ray spectral components that
individually vary in the manner described here would provide evidence
for the interaction between the jet and the accretion-disk radiation
field.  At the earliest times in the flare, the actual disk field is
apparent in the model spectrum at UV, EUV, and soft X-ray energies,
and could be revealed from blazars at UV/soft X-ray energies during
low-intensity states. The direct disk radiation from 3C 279 is argued
to be detected when 3C 279 was faint in the UV \citep{pia99}.

The EC spectral component associated with the isotropic radiation
field would apparently be absent or very weak in lineless or
weakly-lined BL Lac objects, according to general understanding of
these sources.  In the scenario advanced by \citet{bd02} and
\citet{cav02}, BL Lac objects are AGN jet source at a stage in their
life where the fueling is in decline and the black hole engine is most
massive. When GLAST measures BL Lac blazars with better sensitivity,
limits to the $\gamma$-ray flux of a disk radiation component in the
100 MeV - GeV range can be used to infer a relationship between the
location of the acceleration and radiation sites in the jet in terms
of $\delta$, which can itself be inferred from correlated X-ray and
TeV $\gamma$-ray variability \citep{cat97} under the assumption that
the X-rays and TeV $\gamma$ rays have a dominant origin in synchrotron
and SSC processes, respectively.  The near-field and far-field
components have different antennae patterns, with the far-field
component suppressed when observing very close along the jet axis
\citep{dsm92,ds93}.

Fig.\ 4 displays the results of the considered model as a flux density
in units of $\mu$Janskys.  The rapid spectral evolution in the
submillimeter is apparent, and can be tested with submillimeter and
SIRTF observations of the flaring behavior of blazars.  The radio
spectrum is not modeled, and does not properly describe spectra at
frequencies less than the mean synchrotron frequency
$\langle\nu_{syn}\rangle = 2\times 10^6 B\delta\gamma_{min}^2/(1+z)$
Hz radiated by electrons with $\gamma = \gamma_{min}$. The model
moreover breaks down below the synchrotron self-absorption frequency,
which is not treated here. Proper modeling of radio requires a
consistent treatment of adiabatic losses, which is also not done
here. \citet{bot99} points out that the radio spectrum varies
differently in flat spectrum radio sources and BL Lac objects due to
the dominance of the SSC processes in the latter class of sources.

If a $\gamma$-ray flare is a consequence of a relativistic plasma
ejection event whereby the plasma becomes energized, either through an
external or internal shock process as supposed in the model approach
adopted here, then the delayed emergence of a radio-emitting blob is
expected. Evidence for this behavior in EGRET data is presented by
\citet{jor01}. The quasi-continuous monitoring of $\gamma$-ray blazars
with GLAST, in association with VLBI/SIM searches for superluminal
motion in radio jet sources, will reveal whether $\gamma$-ray flares
more typically precede or follow the emergence of radio blobs.

Fig.\ 5 shows the dependence on observing angle $\theta$ of the SEDs,
for the continuous injection model where electrons are steadily
injected during the period that the blob travels from $10^3$ to $10^4
r_g$ (model 1e). Note the increasing dominance of the Thomson
component over the synchrotron component at smaller observing angles
$\theta$ due to the different beaming patterns of the two processes
\citep{der95,dss97,gkm01}. Note also that the period of flaring 
activity for sources observed nearly along the jet is 
smaller than the interval over which the corresponding flux
enhancements are observed at large angles to the jet axis. These
effects are important in statistical treatments of flux-limited
samples of blazar observed at different angles to the observer
direction, which are assumed to be randomly oriented.

\section{Blazar Flare Detectability with {\it GLAST}}

We compare our model with the expected sensitivity of {\it GLAST}, noting
that a recent estimate based on the phase 1 EGRET all sky-survey
\citep{fic94} shows that the rate at which {\it GLAST} will detect
blazar flares sufficiently bright to detect $3\sigma$ variations on 1
hour time scales is about once per month \citep{dd02}.

\subsection{Signal and Background in {\it GLAST}}

The significance of blazar flare detection with {\it GLAST} is
estimated \citep{tho86,dd02}. The number of source photons with
energies $E_1 \leq E < E_2$ detected per unit observer time within the
solid angle element $\Delta \Omega$ centered around a point source in
the direction ($\theta(t),\phi(t)$) with respect to the normal of the
face of the Large Area Telesccope (LAT) tower arrays at time $t$ is
\begin{equation}
\dot S \cong \int_{E_1}^{E_2} dE \; A[E,\theta(t),\phi(t)]\; 
\varphi_s(E,t)\; \{1 - \exp[{-\Delta\Omega\over\Delta\Omega_u(E)}]\}\;,
\label{S}
\end{equation}
where $\varphi_s(E,t)$ is the source photon flux (ph cm$^{-2}$
s$^{-1}$ E$^{-1}$), and $A(E,\theta, \phi)$ is the energy and
angle-dependent effective area of {\it GLAST}. We assume that the
scattered photons are distributed as a Gaussian with $\Delta
\Omega_u(E) = {9\over 4}\pi \theta^2_u(E)$ \citep{tho86}, where
$\theta_u(E)$ is the single photon angular resolution.  This
assumption can be checked against laboratory results to determine the
angular response of the detector and amplitude of non-Gaussian wings
in the point-spread function. The {\it GLAST} requirement for the
single photon angular resolution (68\% containment for on-axis
sources) is $<3.5^\circ$ at 100 MeV and $<0.15^\circ$ at 10 GeV ({\it
GLAST} Science Requirements Document, 2000). Hence $\theta_u =
0.06(E/E_{100})^{-2/3}$, noting the inverse $2/3$ power of the energy
dependence of the point spread function \citep{tho86,ft81}.\footnote{The 
dependence is steeper when $E \lesssim 100$ MeV.}

The source flux $\varphi_s(E,t)$ is related to the $\nu F_\nu$ flux
$f_E($ergs s$^{-1}$) according to the relation $E \varphi_s(E,t) =
f_E/E$. We characterize the $\nu F_\nu$ flux by a power law referred
to 100 MeV (= $E_{100}$) photon energy. Thus
\begin{equation}
f_E \; = \; 10^{-10} f_{-10} ({E\over E_{100}})^{\alpha_\nu} \;{\rm
~ergs~cm}^{-2}{\rm ~s}^{-1}\;,
\label{f_E}
\end{equation}
where $\alpha_\nu$ is the $\nu F_\nu$ spectral index. 

The energy- and angle-dependent effective area of the {\it GLAST}
LAT is approximated by the function
\begin{equation}
A(E,\theta) = A_0 u(\theta)u(\theta)(E/E_{100})^{a(\theta)}\;,
\label{A(E)}
\end{equation}
where $u(\theta = 0^\circ) = 1$, $a(\theta = 0^\circ) = a_0$, and
azimuthal symmetry and time-independence of the detector effective
area is assumed.  An effective area derived from the successful {\it
GLAST} LAT proposal is $A_0 \approx 6200$ cm$^2$ and $a_0 \approx
0.16$ for 100 MeV $\lesssim E \lesssim 10$ GeV. For 20 MeV $\lesssim E
\lesssim 100$ MeV, $a_0 \approx 0.37$. This satisfies the requirement
for on-axis peak effective area of 8000 cm$^2$ in the 1-10 GeV range
({\it GLAST} Science Requirements Document, 2001). Hence the on-axis
effective area $A\cong 6200(E/E_{100})^a$ cm$^2$, where $a = 0.4$, $E<
E_{100}$, and $a = 0.16$, $E>E_{100}$.

Consider a source whose direction $\vec\Omega_s$ is precisely
known. The number of background photons with energies $E$ between
$E_1$ and $E_2$ detected per unit observer time within solid element
$\Delta \Omega$ of the source direction is given by
\begin{equation}
\dot B \cong \Delta \Omega \int_{E_1}^{E_2} dE
\;A[E,\theta(t),\phi(t)]\;\Phi_B(E,\vec\Omega_s)\; .
\label{B}
\end{equation}
The background flux per steradian is denoted by the term
$\Phi_B(E,\vec\Omega)$ (ph cm$^{-2}$ s$^{-1}$ sr$^{-1}$ $E^{-1}$) and
is assumed to be time-independent. Considering only high-latitude
sources where the extragalactic diffuse background radiation dominates
all other sources of background radiation,
\begin{equation}
\Phi_B(E,\vec \Omega) = K_B ({E\over E_{100}})^{-\alpha_B}\; 
\label{Phi_B}
\end{equation}
\citep{sre98} in the range 70 MeV $\lesssim E \lesssim 10$ GeV, where
$K_B = 1.72 (\pm 0.08) \times 10^{-7}$ ph (cm$^2$-s-sr-MeV)$^{-1}$,
and $\alpha_B = 2.10\pm 0.03$.

An exposure factor $X$ is introduced that crudely takes into account
source occultation and variation in detector effective area and photon
localization with changing source direction of {\it GLAST} in its
nominal slewing mode. For flaring behaviors detected on time scales
less than the orbital time scale of 90 minutes ($t_4 =0.54$),
$X$ may approach 1.  On longer time scales, $X \rasim 0.2$. Hence
\begin{equation}
B \cong \Delta\Omega \Delta t X A_0 K_B ({E_1\over E_{100}})^{1+a
-\alpha_B}\cdot 100\; {\rm MeV}\; \cong\; {12\over 1-0.4a}\;X t_4
({E_1\over E_{100}})^{a -2.43}\;,
\label{BE}
\end{equation}
where we let the solid angle acceptance for background $\Delta \Omega
= \Delta \Omega_u$. The approximation in equation
(\ref{BE}) takes into account that most of the background photons are 
collected at the lowest energy of the range of photon energies.

Similar approximations and simplications for the
number of source counts [eq.(\ref{S})] gives the result
\begin{equation}
S\cong 19\;{X t_4 f_{-10}\over 1-{(a+\alpha_\nu)\over 2}}\; 
({E\over E_{100}})^{-2+a +\alpha_\nu }\;.
\label{Sapprox}
\end{equation}
The significance to detect a signal at the $n\sigma$ level for
precisely known background is given by
\begin{equation}
n \cong {S\over \sqrt{B}}
\label{n}
\end{equation}
\citep{lm83}, where $S$ is the number of source counts and $B$ is the
number of background counts.

Equations (\ref{BE}), (\ref{Sapprox}), and (\ref{n}) characterize
high-latitude blazar detectability with GLAST, recognizing also that
detection requires $S \gtrsim$ a few counts. At energies $E\gtrsim
100$ MeV, $a = 0.16$, and we see that
\begin{equation}
{S\over \sqrt{B} }\simeq {5\over 1-{\alpha_\nu\over 2.2}}\; 
\sqrt{Xt_4} f_{-10}\; ({E\over E_{100}})^{\alpha_\nu - 0.71}\;.
\label{SoverrootB}
\end{equation}
Detection of blazar sources is favored at $\approx 100$ MeV photon
energies, except for the hardest sources with $\alpha_\nu \gtrsim 0.7$
(the mean flat spectrum radio quasar photon index in the 100 MeV - 5
GeV range observed with EGRET is $\langle \alpha_\nu\rangle \cong
0.2(\pm 0.2)$
\citep{muk97}). The decline in effective area and the increased
background at lower energies start to hamper source detection
efficiency with {\it GLAST} when $E\ll 100$ MeV.

\subsection{Light Curves}

Light curves of the model blazar flare shown in Figs.\ 2-4 are
plotted in Fig.\ 6, except that here the nonthermal electron
injection ends at $r = 5000 r_g$. Aftter the radiating blob passes
this location, there is no further injection and
the nonthermal electrons cool only through radiative
losses. Thus all three limits specified in the paragraph following
equation (\ref{steady}) are relevant. The absence of adiabatic losses
(see \citet{sik01}), not to mention the assumptions about the
constancy of $B$, $\Gamma$, etc., means that this result illustrates
only a single limit of parameter space. The chief difference
in our model from the recent work by \citet{sik01} is the inclusion of
the direct disk-jet component.

The high-energy light curves at $E = 50$ MeV, 500 MeV, and 5 GeV are
shown in Fig.\ 6a.  The most notable feature is that the 5 GeV light
curve continues to harden while the 50 MeV and 500 MeV light curves
monotonically soften. This is a signature of a disk component fading
as the blob travels outward, while the component from the scattered
disk radiation persists at higher energies. The lower energy light
curves in this figure reach a plateau after injection stops and
electon cooling causes scattered accretion disk radiation to dominate
in this waveband. The photon energy where the two components
contribute equally depends, however, on disk parameters such as mass
accretion rate and inner disk radius, and so might change
for different model parameters.

Fig.\ 6b shows near infrared, X-ray, and medium energy $\gamma$-ray
light curves.  For this model, the synchrotron and SSC components are
equal near 1 keV, and the synchrotron component dominates at lower
energies. Consequently the 0.5 eV IR light curve abruptly declines
after the acceleration episode ends. The X-ray flaring behaviors seen
at 0.5 keV and 5 keV reflect the combined synchrotron and SSC decline
after the injection stops. At 50 keV, the direct accretion-disk
component causes a late time flattening in the light curve, similar to
the behavior seen in the 50 MeV light curve, though the latter is due
to the scattered accretion-disk component. Note also that these light
curves will be smeared out if an integration over light travel time
effects is made \citep{cg99}.

By comparing the blazar flare light curves at $\gamma$-ray energies with
the expression for the significance of detection with {\it GLAST}, given by 
equation (\ref{SoverrootB}), one sees that the flaring behavior will
be detectable, though spectral detail will be difficult to 
extract for the flux given by this model. 
The characteristic signature of a direct accretion
disk component will be indicated in {\it GLAST} data
 by a soft-to-hard behavior at $\sim$ GeV energies
compared to a hard-to-soft behavior at $\sim 100$ MeV energies.  


\section{Summary and Conclusions}

In response to a query from Prof.\ Dr.\ John G. Kirk  about
the applicability of the optically thin approach used in DS93
to an optically-thick Shakura-Sunyaev accretion-disk  spectrum, we have
reexamined the basis and clarified the approach used in the paper by
\citet{ds93} (a different approach is used in the paper by
\citet{dss97}). An optically-thin
formulation is used in the 1993 paper, which appears justified in the
case of accretion-disk models with large radial flow speeds. For
specificity, a photon energy corresponding to the blackbody value was
used, though a wide range of choices may be treated. We conclude that
the method is adequate to treat this special case.

Energy loss rates and spectral functions were derived in the Thomson
regime for the cases of soft photons originating from an external
isotropic monochromatic radiation field, and from a geometrically-thin
accretion disk divided into near-field and far-field regimes in the
limits $\Gamma \gg 1$ and $r \gg \Gamma r_g$ (Appendix A).  SSC
energy-losses were assumed to be small, and the SSC component was
calculated in the Thomson regime. A simplified model was examined, and
a characteristic variability pattern was identified whereby the
near-field accretion disk component, initially bright at $\lesssim 1$
GeV energies, declines in intensity while the component formed by jet
electrons that scatter photons of the external isotropic field becomes
increasingly dominant at $\gtrsim 1$ GeV energies until particle
injection stops. Campaigns organized around {\it GLAST} observations
will be crucially important to interpret the jet-disk interaction by
revealing the physical state of the jet with respect to the disk. The
intensity of the thermal disk during the low state can additionally be
inferred from optical, UV, and X-ray measurements of blazars during
low-intensity states.

Both SSC and EC processes contribute to blazar $\gamma$-ray
production, with a greater contribution of the internal power
dissipated by external Compton (EC) emission in flat radio-spectra
quasars, and a dominant SSC component in BL Lac objects. Multiple
radiation components are required to model well-measured
contemporaneous blazar spectra \citep{bot99,har01,muk99,sik01}. Spectral
variations among different radiation components during different
intensity states has been considered by \citet{bot00}. A
characteristic optical/UV spectrum with $\alpha \cong 3/2$ is found
for injection models with dominating SSC energy losses
\citep{cb02}. The present paper extends this work to provide an
approach to model temporal variations of state transitions, though the
treatment of SSC losses must be improved in future work.

Our model is very preliminary, and we have not even followed a
complete flaring cycle. It is unlikely that the magnetic field remains
constant over thousands or tens of thousands of $r_g$, and analytic models
with $B^2 \propto r^{-2}$ and $B^2 \propto r^{-3}$, or numerical 
models with more general variations of $B$ remain to be 
considered, including the related adiabatic losses. The EC
component from the soft scattered accretion-disk photons will also
decline in intensity as the relativistic ejecta leaves the broad-line
region.  Adiabatic expansion will degrade particle energy until a
dominant synchrotron component remains. The ejecta will expand, cool,
decelerate, and on occasion be reenergized by internal or external
shock interactions. The transition radius from the inner jet, where
the external radiation field is dominated by either direct or
scattered accretion-disk emissions, to the outer jet, where the
external photon field is CMBR dominated, occurs on the kiloparsec
scale for relativistic ejecta.

In the extended jet, the dominant radiation processes are nonthermal
synchrotron radiation, SSC emission, and Compton-scattered CMBR
\citep{hk02}.  Klein-Nishina effects can imprint the spectrum with a
hardening in the Chandra range, as observed in the knots of 3C 273
\citep{mar01,sam01,da02}. As the jet slows to nonrelativistic speeds at a
terminal shock, either synchrotron or SSC process may dominate, as in
the western hot spots of Pictor A \citep{wys01}. Extended jet
formation could be catalyzed by neutral beam production in the inner
jet \citep{ad01,ad02}. The same external-field transformations can be
used to solve the problem of the radiation spectrum from the outer
jet, as presented here for application to the inner blazar jet.

This paper provides a framework to treat many such problems involving
different models and geometries of the accretion-disk radiation field
in both the optically-thick and optically-thin limits, and for other
external radiation fields. A simple extension of these results to
incorporate advective effects on accretion is to change the inner
boundary of the Shakura-Sunyaev disk to $\sim 100 r_g$, within which a
low-luminosity or nonradiating advection-dominated flow is found, with
convection instabilities forming the jet.  Application of these
results to more complicated accretion-disk models, to photomeson
production in blazars, and emission spectra in GRBs will be presented
in subsequent work.

\acknowledgments{We thank John Kirk for his interest in our work, 
and Markus B\"ottcher for discussions. The anonymous referee is 
thanked for a very constructive report.
The work of CD is supported by the Office of Naval Research and the
NASA Astrophysics Theory Program (DPR S-13756G). RS acknowledges
partial support by the Bundesministerium f\"ur Bildung und Forschung
through DESY, grant 05AG9PCA.}

\appendix

\section{Energy Loss Rates in the Near-Field and Far-Field Regimes}

Consider equation (\ref{uprimeext}), 
\begin{equation}
u_{ext}^\prime = {3GM\dot m\over 8\pi c \Gamma r^3} \;
\int_{-\beta_\Gamma}^{\mu^\prime_{max}}d\mu^\prime\; {(\mup +
\bg)^3\over (1-\mu^{\prime 2})^{3/2}(1+\bg\mup )^4}\;= {3GM\dot m\over
8\pi c r^3 \Gamma}(I_{NF}+I_{FF})\;,
\label{uprimeext2}
\end{equation}
valid in the case $\beta_i = 0$, implying an efficiency of 25\% for
this metric.  We have divided the comoving energy density into a NF
component, with $-\bg \leq \mup < 0$, and a FF component with $0 \leq
\mup < \mup_{max}$.

\subsection{Near-Field Integral}

The expansion $\mup = -1 +(N^2/2\Gamma^2)$ in the near-field integral
gives, in the limit $\Gamma \gg 1$,
\begin{equation}
I_{NF} =\int_{-\beta_\Gamma}^0 d\mu^\prime\; {(\mup + \bg)^3\over
(1-\mu^{\prime 2})^{3/2}(1+\bg\mup )^4}\;\;\; \imp \;\;\; 2\Gamma^3
c_d\;,
\label{nfi}
\end{equation}
where
\begin{equation}
c_d \equiv \int_1^\infty dN {(N^2 -1)^3\over N^2(N^2+1)^4} \cong 0.023\;.
\label{nfi2}
\end{equation}
This result agrees with the estimate $c_d \cong 2 \times 3^3/(2^2\cdot
5^4)= 0.0216$, using the value of the integrand at $N = \Gamma\theta
\cong 2$ (the integrand peaks at $\theta^\prime \cong 1.9/\Gamma$ in
the limit $\Gamma \gg 1$).  Fig.\ 7 presents the integrand of equation
(\ref{nfi2}).

\subsection{Far-Field Integral}

The far-field integral is
\begin{equation}
I_{FF} =\int_{0}^{\mu^\prime_{max}}d\mu^\prime\;
{(\mup + \bg)^3\over (1-\mu^{\prime 2})^{3/2}(1+\bg\mup )^4} \;, 
\label{ffi}
\end{equation}
and $\mu_{max} \cong 1 - (R_{min}^2/2r^2)$ in the limit $r \gg
R_{max}$. Hence
\begin{equation}
\mu_{max}^\prime = {\mu_{max} - \bg \over 1-\bg\mu_{max}}\;=\; 
1-2a^2 + O(a^2/\Gamma^2)\;,
\label{mumaxprime}
\end{equation}
in the limit $a \equiv \Gamma R_{min}/r \ll 1$. Therefore
\begin{equation}
I_{FF} \;\;\;\;\iq\;\;\;\; {r\over 4\Gamma R_{min}} \; = \; {1\over 4a}\;.
\label{iffq}
\end{equation}

\subsection{Near-Field/Far-Field Transition Radius}

From equations  (\ref{uprimeext2}), (\ref{nfi}) and (\ref{iffq}), 
\begin{equation}
u_{ext}^\prime = {3GM\dot m\over 8\pi c r^3 \Gamma}
(2\Gamma^3 c_d+{r\over 4\Gamma R_{min}}),
\label{uextprimedef1}
\end{equation}
implying for a Schwarzschild metric with $\beta_i = 0$, the NF/FF
transition radius
\begin{equation}
\tilde r_{NFFF} = \Gamma^4\; ,
\label{rnfff}
\end{equation}
in normalized units $\tilde r = r / r_g$. 

\subsection{Near-Field/Scattered Field Transition Radius}
The location $r_{NX}$ where the NF equals the external scattered
radiation field is given by the condition
\begin{equation}
{3GM\dot m\over 8\pi c r^3 \Gamma}\;\times\; 2\Gamma^3 c_d = 
u^\prime_{NF} = u^\prime_{sc} = \Gamma^2 (1+\bg^2/3)u_* = 
\Gamma^2 (1+\bg^2/3){L_d\tau_{sc}\over 4\pi r_{sc}^2 c}\;.
\label{nfff}
\end{equation}
Recalling equation (\ref{Ldisk2}), we find 
\begin{equation}
\tilde r_{NX} = {r_{NX}\over r_g} =
c_d^{1/3}\,3^{(2+\beta_i)/3}\;{\tilde r_{sc}^{2/3}\over
\tau_{sc}^{1/3}}\cong (0.59 {\rm -} 0.85)\; {\tilde r_{sc}^{2/3}\over
\tau_{sc}^{1/3}}\;,
\label{tildernx}
\end{equation}
where the coefficient varies in value as $\beta_i$ ranges from 0 to 1,
and $r_{sc} = r_g\tilde r_{sc}$ is the characteristic size of the
scattering region, which we identify with the broad-line region.
Hence
\begin{equation}
\tilde r_{NX} \approx  {\tilde r_{sc}^{2/3}\over  \tau_{sc}^{1/3}}\;.
\label{tildernx1}
\end{equation}
Equation (\ref{tildernx1}) holds when $\tilde r_{NX} < \tilde r_{sc}$
for the simplified geometry of the scattering region assumed here,
requiring that $\tilde r_{sc} > 1/\tau_{sc}$. If $r_{sc} = 0.5$ pc
around a $10^9$ Solar mass black hole, then $\tilde r_{sc} =
10^4$. The scattered radiation will dominate the NF radiation beyond
$\approx 2000 (\tau_{sc}/0.01)^{-1/3} r_g$ from the black hole.

\subsection{Far-Field/Scattered Field Transition Radius}

 Equating equations (\ref{dotgammaT}) and (\ref{Edotff}) for the
Thomson energy-loss rates of jet electrons by the quasi-isotropic
scattered disk field and the far field limit of the accretion disk
field, respectively, gives the location beyond which the direct disk
field is no longer important. It is given by
\begin{equation}
\tilde r_{FX} \rightarrow {\sqrt{3}\over 4 \Gamma^2}\;
{\tilde r_{sc}\over \tau_{sc}^{1/2}}
\label{rfx}
\end{equation}
in the limit $\Gamma \gg 1$. Methods of reverberation mapping,
normally applied to nearby radio-quiet Seyferts \citep{pet93}, when
applied to flat spectrum radio quasars with strong lines, such as 3C
273, 3C 279, PKS 0528+134, etc., can be used to infer $\tau_{sc}$ and
the characteristic  size $r_{sc}$ of the scattering clouds. A
simple dependence of $r_{sc}$ on $r_g$ is unlikely, and furthermore,
$r_{sc}$ may depend upon the spectral state and intensity of the
accretion disk.

\subsection{Intensity States of Extragalactic Relativistic Black-Hole
Jet Sources}

These considerations imply a variety of intensity states characterized
by different relative values of
\begin{equation}
\tilde r_{NFFF} = \Gamma^4\; ,\;
\tilde r_{NX} \approx  {\tilde r_{sc}^{2/3}\over  \tau_{sc}^{1/3}}\;
\; , \;  \tilde r_{FX} = { \Gamma^{-2}}\;{\tilde r_{sc}\over \tau_{sc}^{1/2}}.
\label{rcomparison}
\end{equation}
These relations hold within the framework of a geometrically thin,
steady disk that radiates energy through viscous dissipation of
gravitational energy in the absence of advective effects. Both the NF
and the FF spectra are dominant at lower $\gamma$-ray energies than
the external isotropic component. In fact, spectral breaks appear at
$\approx \Gamma^2$ lower energy in the FF than the external isotropic
component.

Flaring behaviors for flat spectrum radio quasars could involve
transitions from the NF to the external scattered field, as modeled in
Figs.\ 2-5. Weaker scattered radiation fields might also involve
transitions from the FF to the external scattered field. In BL Lac
objects, measurements of the intensity of scattered disk components
with GLAST will restrict the distance from the black hole to the jet,
given the specific accretion-disk model. Energy losses due to the SSC
component must then be properly treated to identify the regime where
BL Lac models operate.  Modeling within the context of an ADAF
scenario will improve predictive power and the value of this framework
for the interpretation of multiwaveband observations of relativistic
jet sources. This model is consistent with an evolutionary scenario
where differences between classes of jet sources are a result of
declining dust and gas \citep{bd02}, and will be tested through
statistics of flat spectrum quasars and BL Lac objects \citep{cav02}.

\subsection{Transition from the Inner Jet to the Extended Jet}

A change in spectral state will occur when the pitch-angle averaged
electron Thomson energy-loss rate due to external radiation fields
begins to be dominated by the CMBR rather than the accretion-disk
emission as a consequence of declining intensity of the accretion-disk
field with distance.  This occurs at the radius $r_{exj}$ given by
\begin{equation}
{4\over 3}\Gamma^2 \hat u_{CMB} (1+z)^4 = \chi \; {L_d \over 4\pi
r_{exj}^2 c \cdot 4\Gamma^2}\;,
\label{extjet}
\end{equation}
where $\hat u_{CMB} = 4\times 10^{-13}$ ergs cm$^{-3}$, and $\chi$ an
amplification factor to be explained below.

We write the disk luminosity $L_d = \eta \dot m c^2 = \ell_{\rm Edd}
L_{\rm Edd} = 1.26\times 10^{47}\ell_{\rm Edd}M_9$ ergs s$^{-1}$.
Solving gives
\begin{equation}
r_{exj} = 128 \;{(\chi\ell_{\rm Edd}M_9)^{1/2}\over
\Gamma^2(1+z)^2}\;{\rm~kpc}\; \equiv r_{Xj} \;{\chi^{1/2}\over
\Gamma^2}\;= 320 \;{(\chi\ell_{\rm Edd}M_9)^{1/2}\over
(\Gamma/10)^2[(1+z)/2]^2}\;{\rm~pc}
\label{rexj}
\end{equation}
The radius of the extended jet therefore begins on size scales of
order kpc, though with wide variation depending strongly on $\Gamma$
and less strongly on the accretion-disk luminosity. The quantity $\chi$
represents directional amplification due to the relativistic inner
jet. For the synchrotron component only (the $\gamma$-ray component
would mostly be scattered by jet electrons in the KN limit), it seems
possible that $\chi$ could reach values of 10 or more along the direction
of the inner jet. Deceleration of the ejecta to nonrelativistic speeds
in distant knots and hot spots would make possible the importance of
the external Compton component from the inner jet, as defined by the
quantity $r_{Xj}$. Thus the disk/inner jet radiation field could be
important in spectral models of hot spots $\sim 10^2$ kpc from the
central black hole when the accretion-disk or inner-jet power flares
to $\sim L_{\rm Edd}$ over long time scales ($\gg 10^3$ yrs).

\section{Scattered Accretion-Disk Radiation Spectrum in the Near-Field Regime}

From the analysis of the integrand of equation (\ref{nfi}) in the NF
regime, we see that the NF photons originate from $\theta^\prime \cong
1.9/\Gamma$, or from $r \approx R$ in the stationary frame. From
equations (\ref{Ldisk2}), (\ref{uprimeext2}) and (\ref{nfi}), the
comoving energy density in the NF regime is
\begin{equation}
u^{\prime~0}_{NF} = {L_d\over 4\pi r^2 c}\;4\cdot 3^{1+\beta_i} \; c_d
({r_g\over r})\Gamma^2\;,
\label{nfu}
\end{equation}
and $4\cdot\ 3^{1+\beta_i} \; c_d = 0.28, 0.83$ when $\beta_i = 0,1$,
respectively. Taking the peak contribution from the disk at $\mu =
1/2$, we have the following approximation for the comoving energy
density:
\begin{equation}
u^{\prime}_{NF} = {u^{\prime~0}_{NF}\over 2\pi}\; \delta(\ep -
\Gamma\e_*/2)\delta(\mup+1)\;.
\label{uprimeapnf}
\end{equation}

The comoving emissivity can be derived from 
\begin{equation}
j^\prime(\ep_s,\Omega^\prime_s) = c \int_0^\infty d \ep\oint
d\Omega^\prime \int_1^\infty d\gamma\oint
d\Omega_e\;(1-\beta\cos\psi)({\ep_s\over \ep})
u_{ph}^\prime(\ep,\Omega^\prime )n_e(\gamma,\Omega_e)\;({d\sigma\over
d\ep_s d\Omega_s})\;
\label{jprime}
\end{equation}
\citep{dss97}, and $\cos\psi
=\mu_e\mup+(1-\mu_e^2)^{1/2}(1-\mu^{\prime~2})^{1/2}\cos(\phi_e
-\phi^\prime)$. Equation (\ref{jprime}) allows to examine the
evolution of the particle distribution function in the comoving
frame. An important simplifying assumption we make at this point is to
assume isotropy of the electron distribution function in the comoving
frame, that is,
\begin{equation}
n_e(\gamma,\Omega_e) = {n_e(\gamma)\over 4\pi}\;.
\label{n_e}
\end{equation}
For the cross section in the Thomson regime, we make the approximation
\begin{equation}
{d\sigma\over d\ep_s d\Op_s} = \sigma_{\rm T}
\delta[\ep_s-\gamma^2\ep(1-\beta\cos\psi)]\delta(\Omega^\prime_s
-\Omega_e)\;.
\label{dsig}
\end{equation}
Solving gives
\begin{equation}
j^\prime(\ep_s,\Op_s)= {c\sigma_{\rm T}u^{\prime~0}_{NF}\over
2\pi}\;{(1+\mup_s)^2\over 4\ep_s}\;\gamma_{NF}^3
n_e(\gamma_{NF})\;\;,\;\;\gamma_{NF} = \sqrt{{2\ep_s\over
\Gamma\e_*(1+\mup_s)}}\;.
\label{jprimeNF}
\end{equation}
The $\nu F_\nu$ spectrum 
\begin{equation}
f_\e^{NF}={\delta^4\over d_L^2}\; V_b \ep_s
j^\prime(\ep_s,\Omega^\prime_s)\;,\;\delta =
[\Gamma(1-\bg\mu_s)]^{-1}\;,\;.
\label{nuFnub}
\end{equation}
 Making the transformation to observer frame quantities using the 
relation $1+\mu^\prime_s \rightarrow \delta(1+\mu_s)/2\Gamma$, valid when
$\Gamma \gg 1$, gives
\begin{equation}
f_\e^{NF}={\delta^6 c\sigma_{\rm T} u^{\prime ~0}_{NF}\over 32\pi
d_L^2\Gamma^2}\;(1+\mu_s)^2\hat\gamma^3 N_e(\hat\gamma)\;,\;\hat\gamma
= {2\over \delta}\;\sqrt{{(1+z)\e\over \e_* (1+\mu_s)}}\;,
\label{nuFnub1}
\end{equation}
from which equations (\ref{Edotnf}) and (\ref{fnf}) follow. A value
intermediate to 0.28 and 0.83 is assigned in equations (\ref{Edotnf})
and (\ref{fnf}).

\section{Scattered Accretion-Disk Radiation Spectrum in the Far-Field Regime}

The point-source approximation for a monochromatic radiation field
coming directly from behind the direction of jet motion is
\begin{equation}
u_{FF} (\e,\Omega) = {L_d\over 4\pi r^2 c}\; {\delta (\mu - 1)\over 2
\pi }\; \delta(\e - \bar\epsilon )\;.
\label{uphb}
\end{equation}
Making use of the relations $\e = \Gamma \ep (1+\bg\mup)$, $\mu =
(\mup+\bg)/(1+\bg\mup)$, and $\phi = \phi^\prime$ gives
\begin{equation}
u_{FF}^\prime (\ep,\Omega^\prime) = {L_d\over 8\pi^2 r^2 c}\; {\delta
(\mup - 1)\delta[\ep - \bar\epsilon/\Gamma(1+\bg)]\over
\Gamma^2(1+\bg^2) }
\label{uphbprime}
\end{equation}

Following the same procedure as in Appendix B gives
\begin{equation}
j^\prime(\ep_s,\Omega^\prime_s) = {\sigma_{\rm T}L_d (1-\mup_s)\over
32 \pi^2 r^2\bar \epsilon\Gamma(1+\bg)}\;\hat\gamma
n_e(\hat\gamma)\;,\;\hat\gamma= [{\Gamma\ep_s(1+\bg)\over \bar\epsilon
(1-\mup_s)}]^{1/2}\;,
\label{jprime2}
\end{equation}
so that
\begin{equation}
f_\e^{FF} = \delta^6\; {c\sigma_{\rm T}(1-\mu_s^2)\over 8\pi
d_L^2}\;({L_d\over 4\pi r^2
c})\;\tilde\gamma^3N_e(\tilde\gamma)\;,\;\tilde\gamma \equiv
\delta^{-1}\sqrt{{(1+z)\e_s\over (1-\mu_s)\bar\epsilon}}\;,
\label{feb}
\end{equation}
giving equation (\ref{fff}).

The accuracy of the expression for $j^\prime(\ep_s,\Omega^\prime_s)$
in equation (\ref{jprime2}) can be checked by deriving the comoving
electron energy-loss rate
\begin{equation}
-m_ec^2\dot\gamma = \oint d\Op_s \int_0^\infty\;d\ep_s\;V^\prime_b
j^\prime(\ep_s,\Omega^\prime_s)\;\;,\;{\rm when}\;\;
V^\prime_b n_e(\hat\gamma) = \delta(\hat\gamma -\bar \gamma)\;,
\label{gdotb}
\end{equation}
from which follows equation (\ref{Edotff}).

\section{Klein-Nishina Effects in Compton Scattering}

\citet{gkm01} have derived an expression 
for the radiation spectrum produced by an isotropic distribution of
electrons that Compton-scatters external soft photons with
monochromatic dimensionless energy $\e_*$ and energy density $u_*$ in
the stationary frame.  Slightly modified, their result is
\begin{equation}
f_\e^{\rm KN} = \delta^3\;{3\sigma_{\rm T} u_*\over 16\pi
d_L^2}\;[{\e(1+z)\over \e_*}]^2 \;
\int_1^\infty d\gamma_* \;{F(x) N(\gamma_*/\delta)\over \gamma_*^2}\;
H[\gamma_*;\delta\gamma_1,\delta\gamma_2]\;,
\label{feKN}
\end{equation}
where
\begin{equation}
F(x) = [2x\ln x + x - 1 - 2x^2 + {(4\e_*\gamma_*x)^2\over
2(1+4\e_*\gamma_* x)}]\;H[x;{1\over 4\gamma_*^2}, 1]\;\;,\;\;{\rm and}
\;\;x = {\e(1+z)\over 4\gamma_*^2 \e_*}\;
\label{F(x)}
\end{equation}
This expression employs the head-on approximation for the Compton
scattering process \citep{jon68,bg70}. The comoving electron spectrum
in the range $\gamma_1 \leq \gamma \leq \gamma_2$ is given by
$N(\gamma)$. Klein-Nishina effects become important when $\epsilon
\gtrsim 1/[(1+z)\e_*]$.

For a power-law electron distribution $N(\gamma) = V_b^\prime
k_e\gamma^{-p}H[\gamma;\gamma_1,\gamma_2]$, where $V^\prime_b$ is the
comoving blob volume, they also derive an accurate form for the
Thomson regime expression of the specific spectral power, given by
\begin{eqnarray}
L_{\rm GKM}(\e,\Omega) = \delta^{3+p}\;{V_b^\prime k_e c\sigma_{\rm T}
u_*\over 8\pi\e_*}\;({\e\over\e_*})\;
\end{eqnarray}
\begin{equation}
\times\{ (\gamma_2\delta)^{-(1+p)}
[{\e\over 4\e_*(p+3)(\gamma_2\delta)^2}-{1\over 1+p}]+
({\e\over 4\e_*})^{-(1+p)/2}
\;{2\over (1+p)(3+p)}\} \;,
\label{LGKM}
\end{equation}
and show that it reduces to the expression
\begin{equation}
L_{x T}(\e,\Omega) = \delta^{p+3}\;{c\sigma_{\rm T} u_*\over \pi
(p+3)}\;V^\prime_b k_e\;2^{(p-3)/2}\;({\e\over \e_*})^{(1-p)/2}\;
\label{LDSS}
\end{equation}
when $p > -1$, $\e \ll 4\delta^2\e_*\gamma_2^2$, and $\gamma_2\gg
\gamma_1$.  The Thomson scattered $\nu F_\nu$ spectrum of an external
isotropic and monochromatic radiation field was derived by
\citet{dss97} and is given by
\begin{equation}
f_\e^{xT} = {\delta^{4+2\alpha}\over {\alpha+2}}\;{c\sigma_{\rm T}
u_*\over 16\pi d_L^2} \;
k_eV_b^\prime(1+z)^{1-\alpha}\;{(1+\mu)^{2+\alpha}\over
\mu}\;({\e\over\e_*})^{1-\alpha}\;,
\label{fext}
\end{equation}
where $\alpha=(p-1)/2$. Equation (\ref{fext}) recovers equation
(\ref{LDSS}) in the limit $\mu \rightarrow 1$, though the term
$2^{(p-3)/2} = 0.71, 0.84,$ and 1.0 for $p = 2,2.5,$ and 3,
respectively, in equation (\ref{fext}), is replaced by $2^{p-1}/(p+1)
= 0.67, 0.81,$ and 1.0 for $p = 2,2.5,$ and 3, respectively, in
equation (\ref{LDSS}). The factor $(1+\mu)^{(2+\alpha)}/\mu$ follows
from the range of scattered photon energies and angles in the comoving
frame, and indicates that $\mu > 0$ is the range of validity of the
expression.

The approach of \citet{gkm01} avoids the need to transform scattered
radiation spectra from the comoving to observer frames by directly
transforming electron energy spectra to the stationary frame. This is
useful in the case of specified comoving electron distributions. When
calculating evolving electron spectra, the external radiation fields
must be transformed to the comoving frame in order to derive the
electron cooling rates. Furthermore, if the electron distribution is
not assumed to be isotropic in the comoving frame, it is also simpler
to treat the evolving energy and angle-dependent transformations of
the electron spectra in the comoving frame. Whether one performs the
scattering in the comoving frame and then transforms the radiation
spectrum, or transforms the electron spectrum to the stationary frame
and then performs the scattering, the final result must of course be
the same.

\citet{gkm01} showed that the Thomson approximations
(\ref{LGKM})-(\ref{fext}) greatly overestimates the scattered
radiation spectra compared to spectra calculated using equation
(\ref{feKN}). This discrepancy is considerably reduced for electron
spectra that evolve in response to Thomson losses in the
Thomson-scattering approximation, compared with electrons that evolve
under self-consistent Compton losses when using equation (\ref{feKN}).
An approximate analytic treatment of KN losses and a detailed
numerical treatment of joint Thomson and KN energy losses resulting
from the CMBR field is performed by \citet{da02}, and \citet{bms97}
calculate blazar spectra using the full Compton cross section.

\clearpage

\begin{figure}
\vskip-2.0in
\plotone{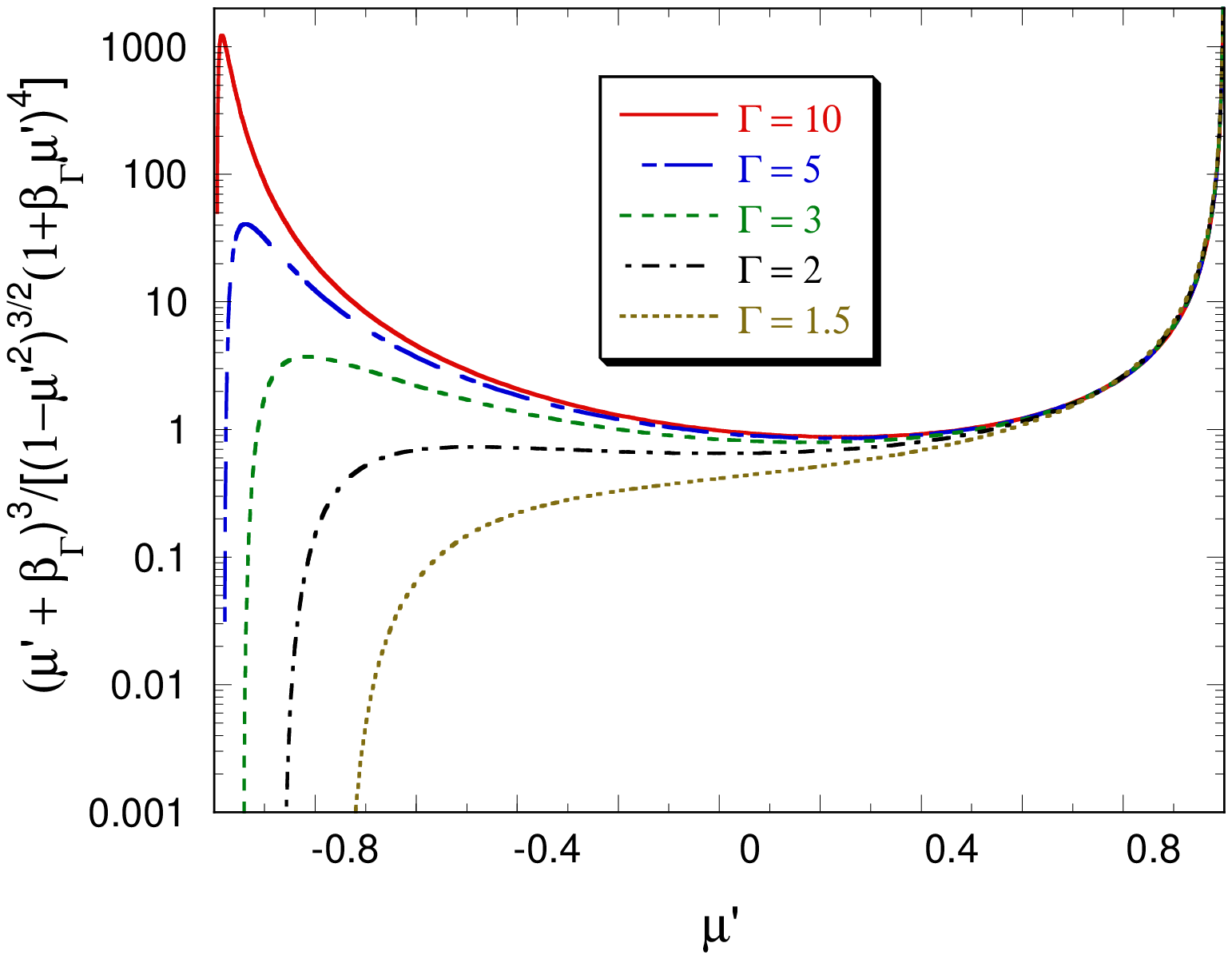}
\caption{Integrand of integral on right-hand-side 
of equation (\ref{uprimeext}).}
\label{f1}
\end{figure}

\clearpage
\begin{figure}
\vskip-2.0in
\plottwo{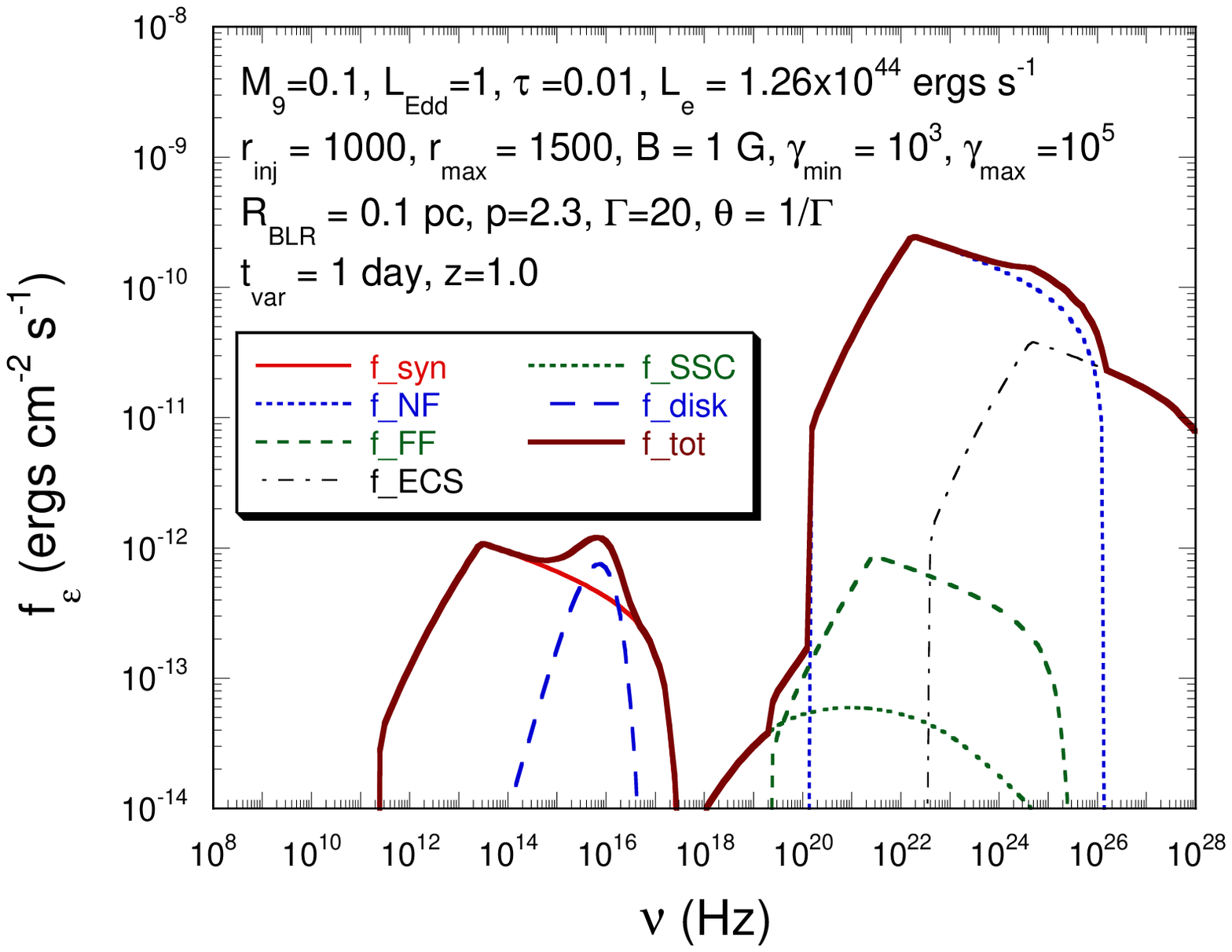}{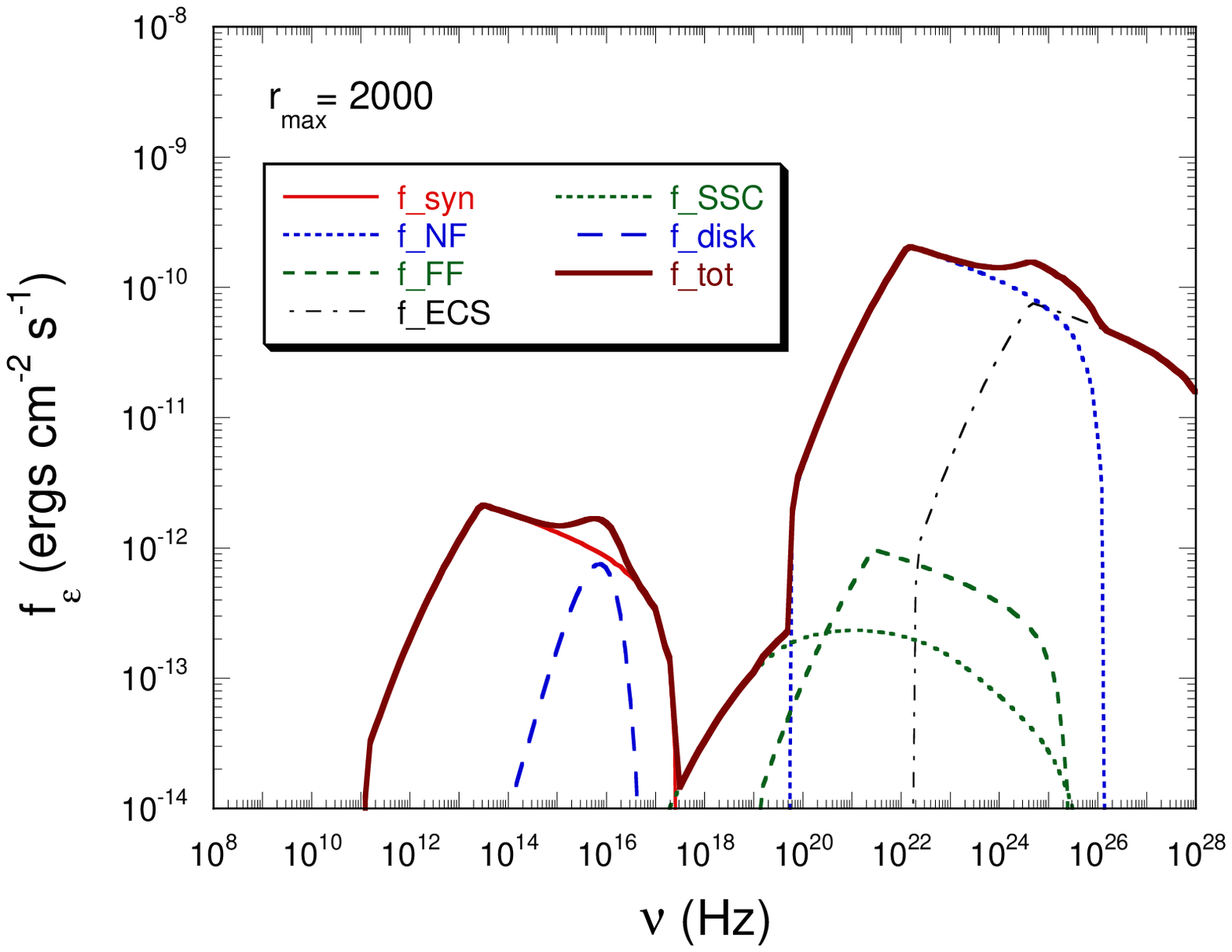}
\end{figure}
\begin{figure}
\vskip-2.0in
\plottwo{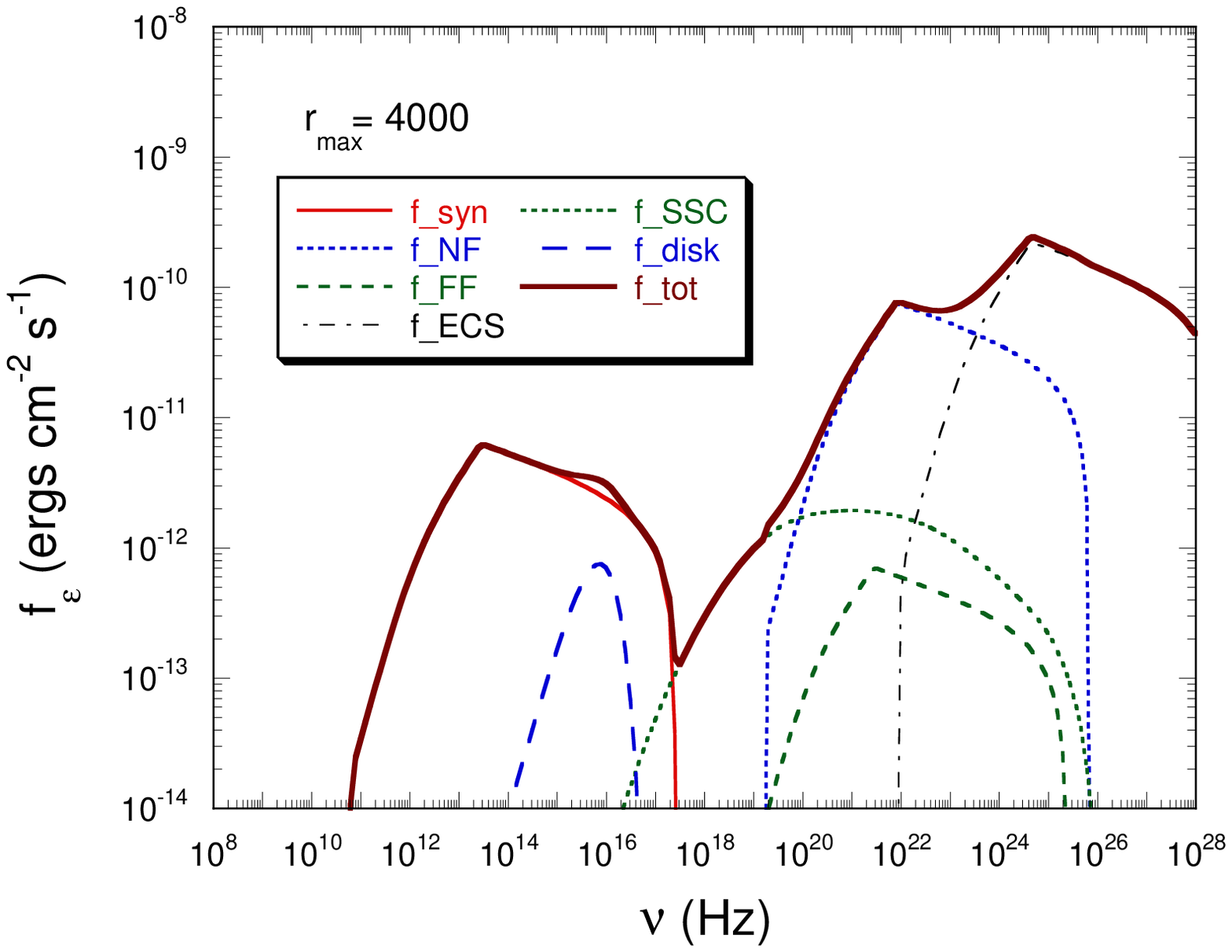}{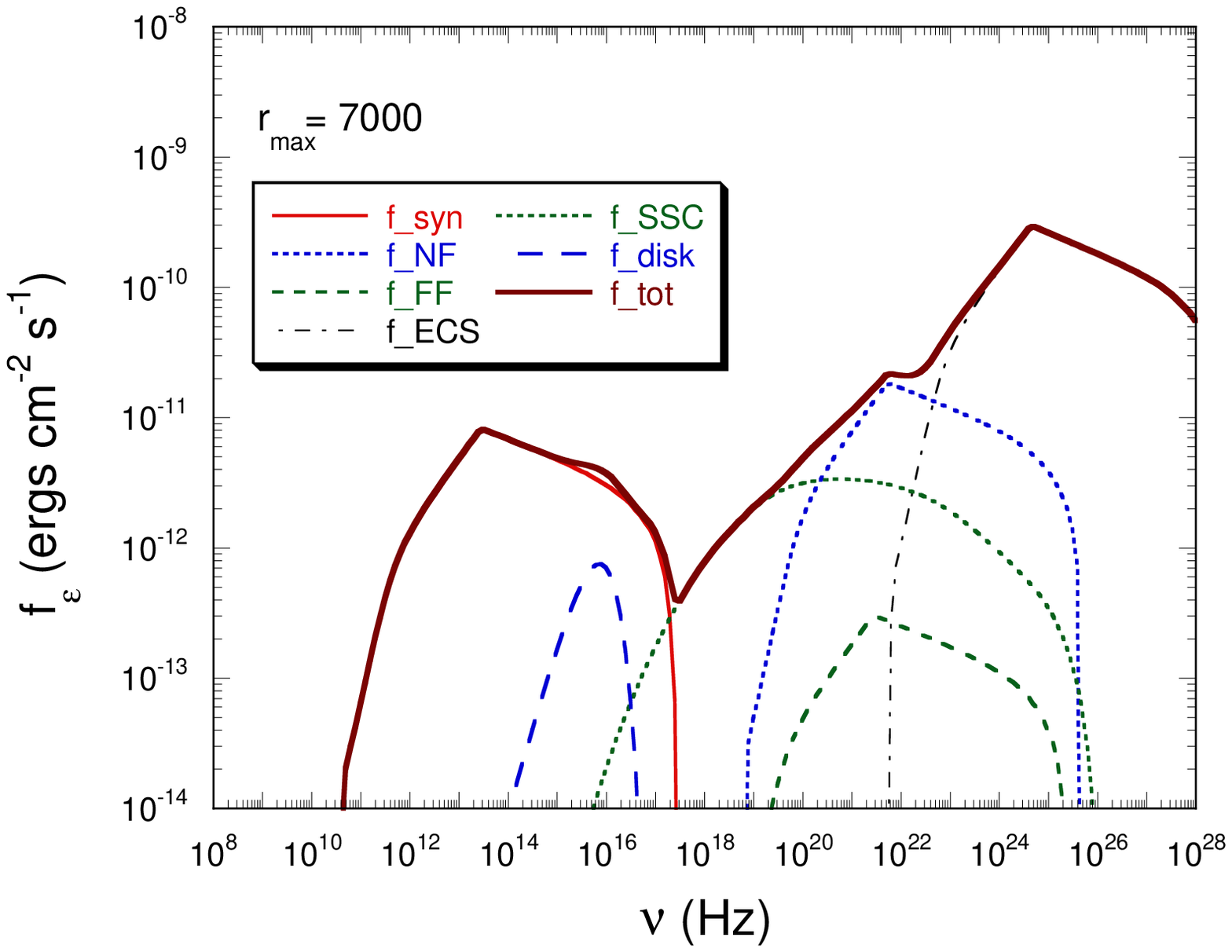}
\end{figure}

\begin{figure}
\epsscale{0.7}
\vskip-2.0in
\plotone{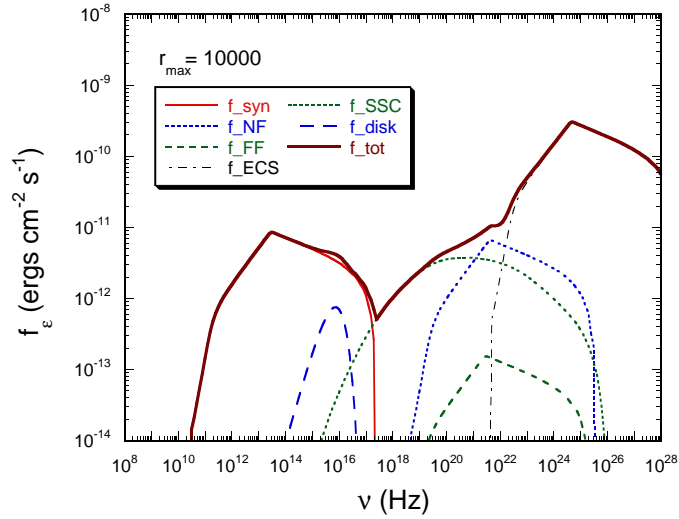}
\caption{Spectral model of blazar variability, 
including an SSC component, and EC
contributions from the direct near-field and far-field disk field 
and the quasi-isotropic
scattered radiation field. Parameters and identification of 
the separate spectral components 
are given in the legend. (a) Electrons are uniformly injected 
with a comoving power of $1.26\times 10^{44}$ 
ergs s$^{-1}$ as 
the jet travels from $r = 10^3 r_g$ to $r = 1500 r_g$.
(b) Same as in Fig.\ 2a, but with the electrons 
uniformly injected between $r = 10^3 r_g$ and $r = 2000 r_g$.(c)
 Same as in Fig.\ 2a, but with electrons 
uniformly injected between $r = 10^3 r_g$ and $r = 4000 r_g$. (d) 
Same as in Fig.\ 2a, but with electrons 
uniformly injected between $r = 10^3 r_g$ and $r = 7000 r_g$.(e)
Same as in Fig.\ 2a, but with electrons 
uniformly injected between $r = 10^3 r_g$ and $r =10^4 r_g$.
}
\label{f2a}
\end{figure}

\begin{figure}
\epsscale{1.0}
\vskip-2.0in
\plotone{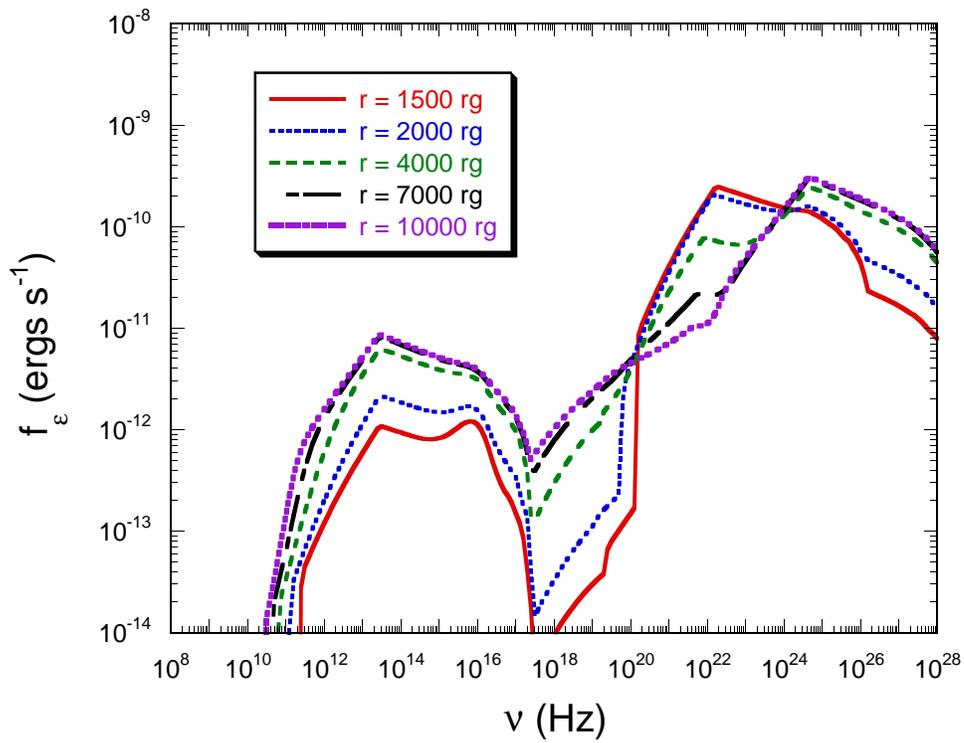}
\caption{Multiwavelength variability behavior 
for a uniform injection event for the model in Fig.\ 2, beginning
when the jet is at $r = 10^3 r_g$, 
and ending when $r = 10^4 r_g$. Note the decline of the 
direct disk-radiation field component with distance from the black hole. }
\label{f3}
\end{figure}

\begin{figure}
\vskip-2.0in
\plotone{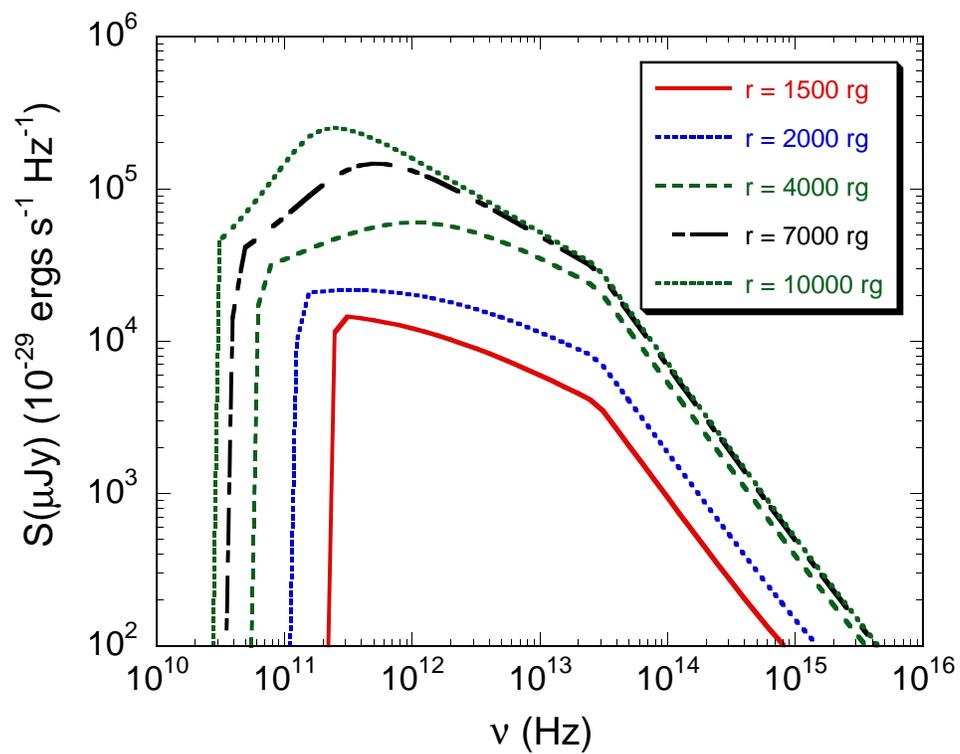}
\caption{Flare profile presented as a 
flux density in units of micro-Janskys, showing
the characteristic submillimeter/infrared behavior 
for this class of flare. The radio spectrum is
not modeled here.}
\label{f4}
\end{figure}

\begin{figure}
\vskip-2.0in
\plotone{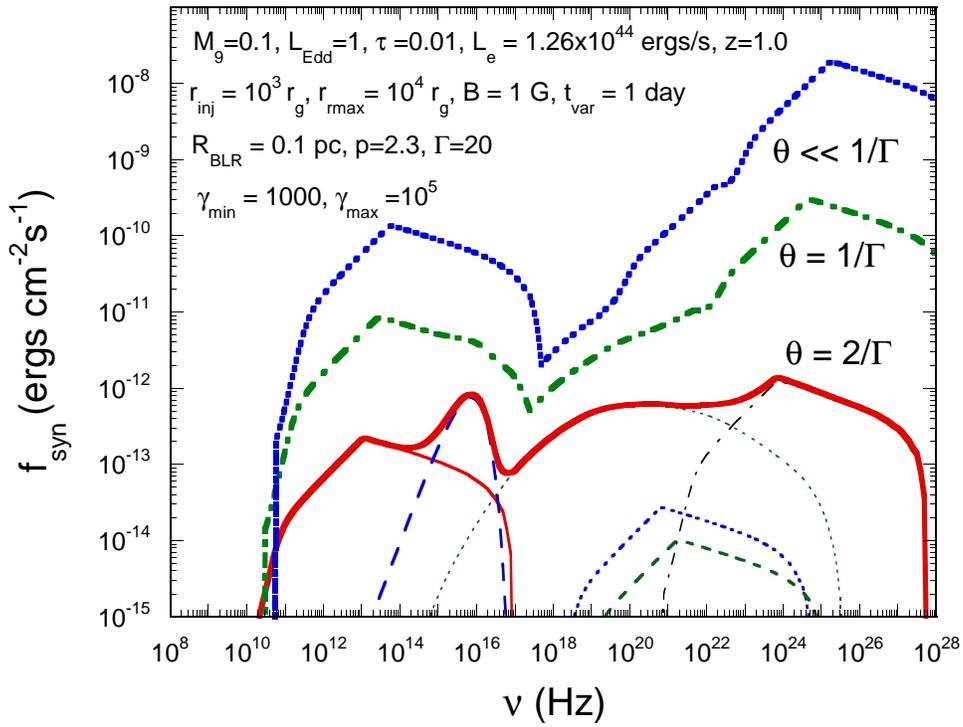}
\caption{Angle-dependences of received 
$\nu F_\nu$ flux toy model blazar flare. 
The dominant Compton component at small angles
 reflects the different beaming
factors of synchrotron, SSC and EC processes.}
\label{f5}
\end{figure}

\clearpage
\begin{figure}
\vskip-2.0in
\plottwo{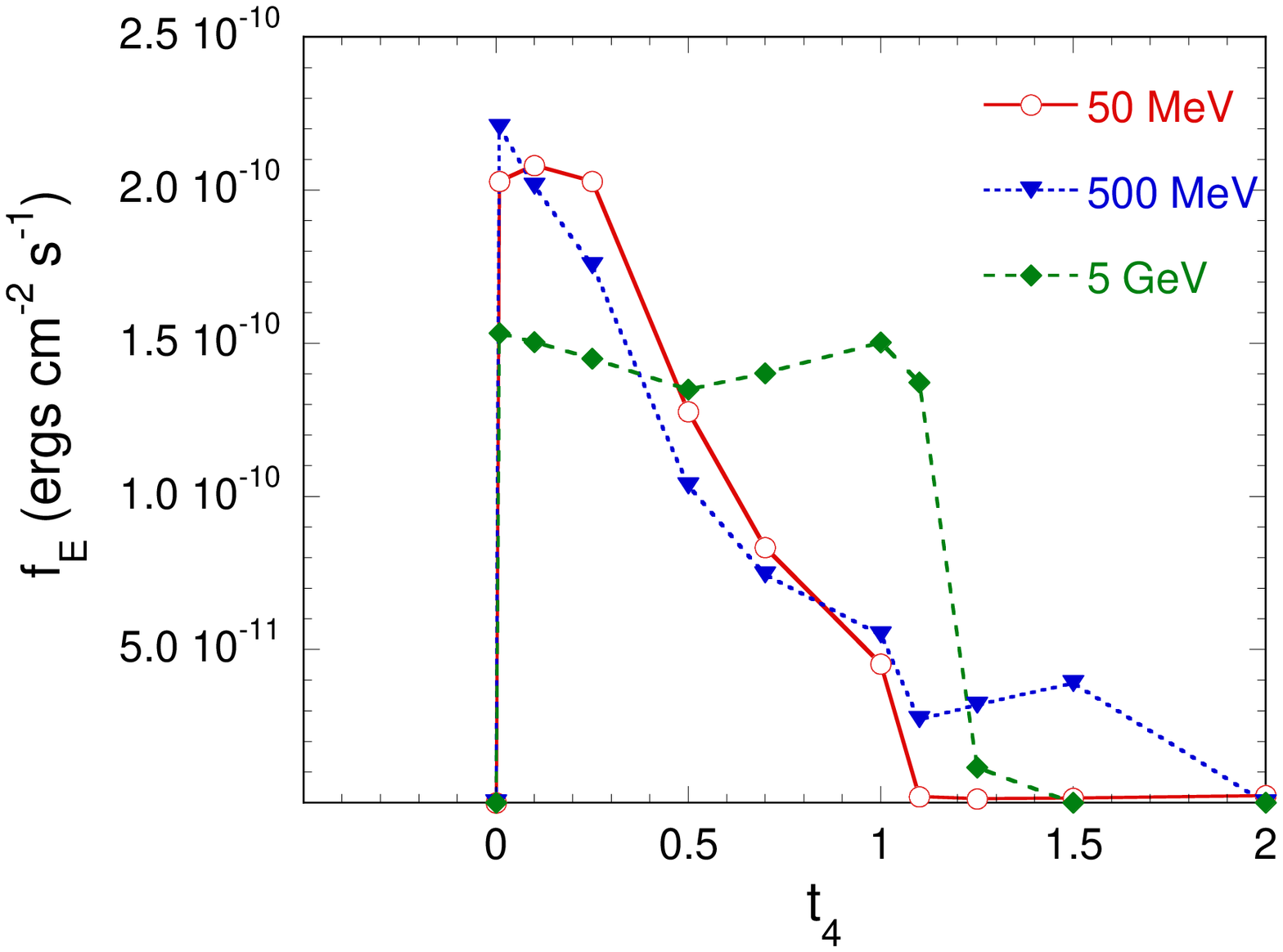}{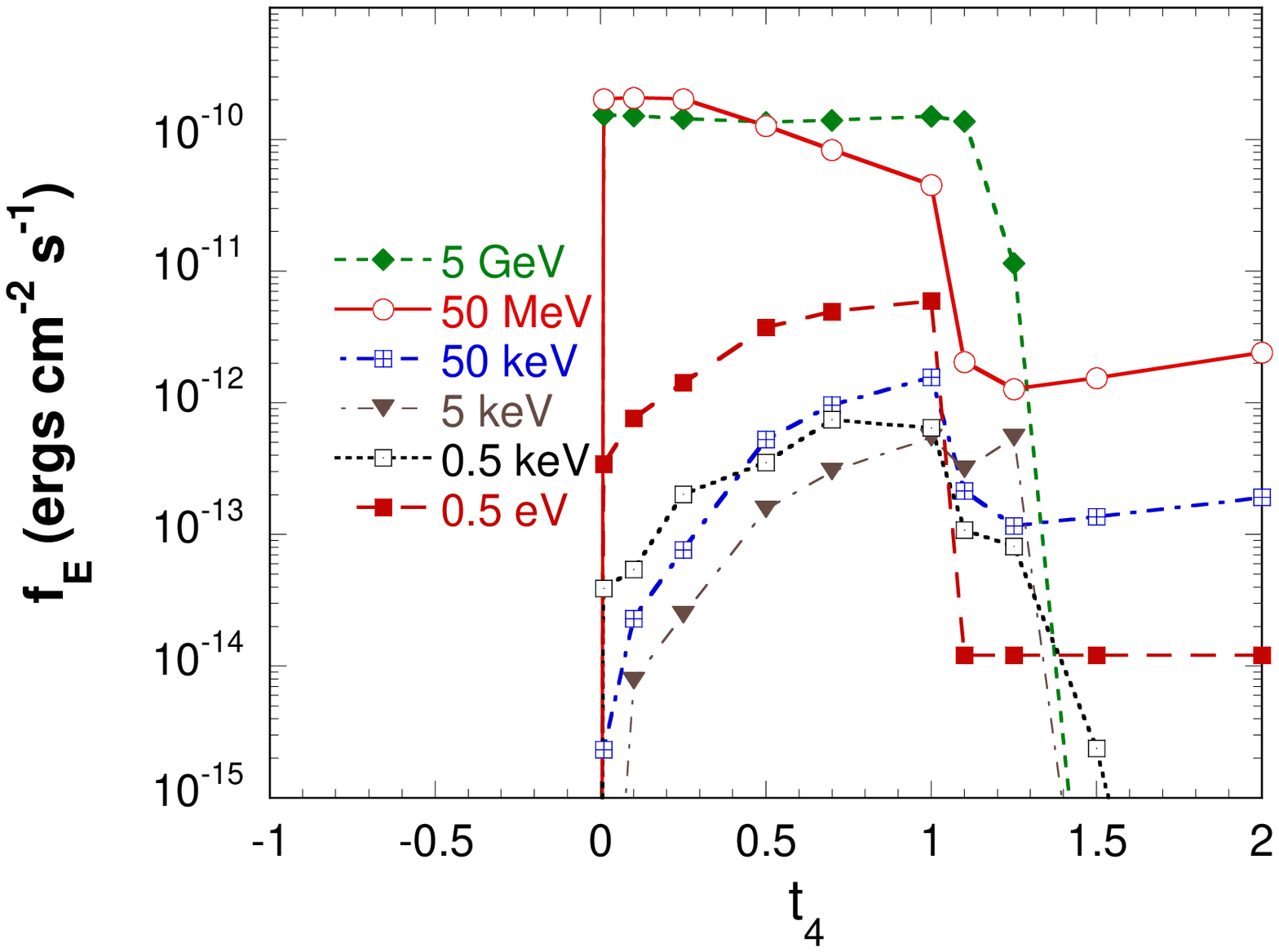}
\caption{Light curves associated with a model flare where
injection begins at $r = 10^3 r_g$ and ends at $r = 5\times 10^3 r_g$.
Except for the time when particle injection stops, all other parameters
are the same as in Fig.\ 2. Time is measured from the start of the flare
in units of $10^4$ s. (a) Light curves at 50 MeV, 500 MeV, and 5 GeV.
(b) Light curves at 0.5 eV, 0.5 keV, 5 keV, 50 keV, 50 MeV, and 5 GeV.}
\label{f6}
\end{figure}

\begin{figure}
\vskip-2.0in
\plotone{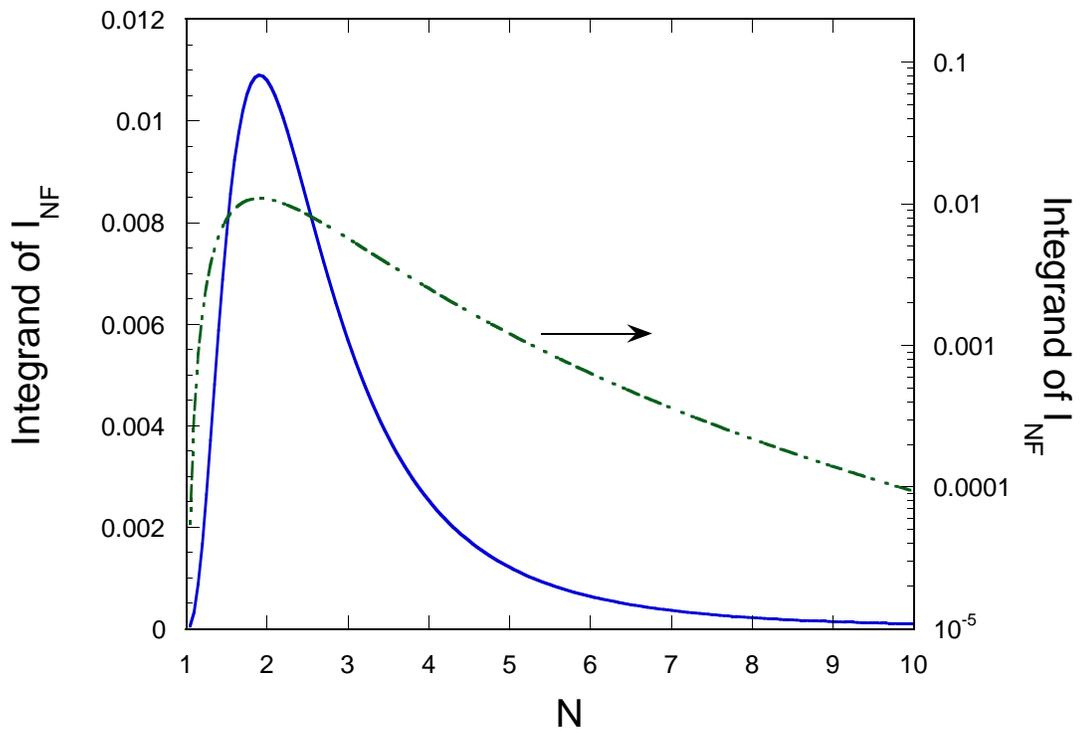}
\caption{The integrand $(N^2 -1)^3/[ N^2(N^2+1)^4]$ 
of equation (\ref{nfi2}). }
\label{f7}
\end{figure}

\end{document}